\documentclass[11pt]{article}
\usepackage{graphicx}
\usepackage{amsmath,amssymb,amsfonts,amsthm}
\usepackage{setspace}
\usepackage{subfig}
\usepackage{booktabs}
\usepackage{multirow}
\usepackage{hyperref}
\usepackage{float}
\usepackage{color}
\usepackage{natbib}
\usepackage[normalem]{ulem}
\useunder{\uline}{\ul}{}
\usepackage{booktabs}
\usepackage{adjustbox}
\usepackage{tikz}
\usepackage{comment}
\usepackage{booktabs,siunitx,multirow,adjustbox}
\hypersetup{
	colorlinks,%
	citecolor=black,%
	filecolor=black,%
	linkcolor=black,%
	urlcolor=blue
}
\usepackage[top=1 in, bottom=1 in, left=1 in , right=1 in]{geometry}
\usepackage{titlesec}
\titleformat{\section}{\Large\bfseries}{\thesection}{1em}{}

\usepackage{enumitem}
\setdescription{leftmargin=\parindent,labelindent=\parindent}

\begin{document}
\author{$%
	\begin{array}
		[c]{ccc}%
		\text{Eunkyu Seong}\thanks{Department of Economics, Seoul National University. Email: uocup96@snu.ac.kr}& \hspace{0.2in} & \text{Seojeong Lee}\thanks{Department of Economics, Seoul National University. Corresponding Author. Email: s.jay.lee@snu.ac.kr} \ \thanks{Lee acknowledges that this work was supported by the Research Grant of the Center for National Competitiveness at the Institute of Economic Research, Seoul National University.}
%		\text{Seoul National University} & \hspace{0.2in} & \text{Seoul National University}\\
	\end{array}
	\medskip$}
\date{\today}

\title{Are the Bank of Korea's Inflation Forecasts\\ Biased Toward the Target?}
%\date{}
\maketitle

\begin{abstract}
The Bank of Korea (BoK) regularly publishes the \textit{Economic Outlook}, offering forecasts for key macroeconomic variables such as GDP growth, inflation, and unemployment rates. This study examines whether the BoK’s inflation forecasts exhibit bias, specifically a tendency to align with its inflation target. We extend the Holden and Peel (1990) test to incorporate state-dependency, defining the state of the economy based on whether realized inflation falls below the target at the time of the forecast. Our analysis reveals that the BoK’s inflation forecasts are biased under this state-dependent framework. Furthermore, we examine a range of bias correction strategies based on AR(1) and mean error models, including their state-dependent variants. These strategies generally improve forecast accuracy. Among them, the AR(1)-based correction exhibits relatively stable performance, consistently reducing the root mean square error.
\end{abstract}


%This research was supported by the BK21 FOUR (Fostering Outstanding Universities for Research) funded by the Ministry of Education(MOE, Korea) and National Research Foundation of Korea(NRF). 

\section{Introduction}

The Bank of Korea (BoK) regularly publishes the \textit{Economic Outlook}, providing forecasts for key macroeconomic indicators such as GDP growth, inflation, and unemployment rates. These forecasts are crucial for informing government policies and guiding firms’ investment decisions. Despite their significance, limited research has rigorously evaluated the accuracy and potential biases of these forecasts. This paper addresses this gap by conducting a comprehensive ex post evaluation of the BoK’s economic forecasts, comparing their projections to actual realized values.

Our findings reveal a notable bias in the BoK’s inflation forecasts, which tend to align with the institution’s inflation target. In contrast, forecasts for GDP growth and unemployment rates do not exhibit similar biases. To formally investigate these patterns, we extend the traditional \citet{holden1990testing} test to account for state-dependency, defining the state of the economy based on whether the actual inflation rate falls below the target at the time of the forecast.

The results of our state-dependent tests reject the null hypothesis of unbiased inflation forecasts, irrespective of the economic state. This contrasts with prior studies, which generally find the BoK’s inflation forecasts to be unbiased and efficient (\citet{kpl2011}). Specifically, we find that one-quarter ($h=1$) and three-quarters ahead ($h=3$) inflation forecasts tend to upward biased when inflation is below the target and downward biased when it is above.

Beyond identifying forecast bias, we explore potential bias-correction strategies using only the BoK’s forecasts and realized values. Our analysis demonstrates that an autoregressive (AR(1)) bias correction strategy improves forecast accuracy, as evidenced by a reduction in root mean square forecast error (RMSFE).

%In this paper we investigate the Bank of Korea's economic projections on various macro variables, and especially inflation rate. We collect halfyearly data from 1999h1 to 2023h2 from the quarterly report `Economic Forecast'. We discover that the Bank of Korea's Inflation forecasts are biased and this bias tends toward the inflation target. We also compare several potential strategies of economic agents how to response to BoK's announcement, and show that there exists a strategy which dominates "naive" strategy, in RMSE-sense.

%In section 2 we describe the structure of data in detail. In section 3 we test for the unbiasedness and efficiency of BoK's forecasts. In section 4 we present simulation results of several alternative strategies. In section 5 we summarize the results and conclude.

\section{Literature}

Numerous studies have evaluated the accuracy, unbiasedness, and efficiency of economic forecasts produced by various institutions worldwide. For example, several studies examine the forecasts published in the World Economic Outlook (WEO) by the International Monetary Fund (IMF), including \citet{timmermann2007evaluation}, \citet{aldenhoff2007economic}, and \citet{koch2024we}. \citet{timmermann2007evaluation} derives testable implications based on the assumption that forecasters have symmetric loss functions and behave rationally, suggesting that forecast errors should have an expected value of zero and that forecast errors and revisions should not be predictable. \citet{aldenhoff2007economic} investigates whether WEO forecast bias correlates with election periods, while \citet{koch2024we} examine channels of failure in inflation forecasting during the post-COVID-19 period.

Studies focusing on the Federal Reserve’s forecasts include \citet{scotese1994forecast}, \citet{romer2000federal}, \citet{capistran2008bias}, and \citet{sheng2015evaluating}. \citet{scotese1994forecast} analyzes the Federal Reserve staff’s forecasts for real GNP and inflation, testing for the “reputation effect,” which incentivizes forecasters to smooth predictions. Similarly, \citet{romer2000federal} use Greenbook GNP deflator forecasts to examine the Fed’s information efficiency and whether Greenbook forecasts incorporate the information of other professional forecasts. \citet{capistran2008bias} uses Greenbook data to recover the Fed’s loss function, revealing that overestimating inflation is less costly than underestimating it. \citet{sheng2015evaluating} analyzes economic forecasts by Federal Open Market Committee (FOMC) members, attributing forecast heterogeneity partly to regional economic conditions and partly to individual preferences for monetary policy.

The European Central Bank (ECB) forecasts have also been the subject of study, as seen in \citet{alessi2014central}, \citet{granziera2025bias}, \citet{kontogeorgos2022evaluating}, and \citet{argiri2024evaluation}. \citet{alessi2014central} examine the performance of point forecasts by the ECB and the Federal Reserve Bank of New York, focusing on the integration of financial market signals during the global financial crisis. \citet{kontogeorgos2022evaluating} use confidential real-time ECB forecast data to assess forecast unbiasedness and efficiency.

Other central banks have also been studied. \citet{charemza2016central} develop a model showing that inflation forecasts by central banks in inflation-targeting countries tend to bias toward the target. \citet{knuppel2019assessing} evaluate inflation uncertainty forecasts from the Bank of England, Central Bank of Brazil, National Bank of Hungary, and Sveriges Riksbank, finding that short-horizon forecasts are underconfident, while long-horizon forecasts are overconfident.

As forecast evaluation literature has grown, methodological advancements have been made. \citet{komunjer2012multivariate} introduce hypothesis-testing methods for vectors of forecast errors, applied to Federal Reserve forecasts by \citet{caunedo2020asymmetry} and \citet{sinclair2015evaluating}. \citet{rossi2016forecast} propose a framework for forecast rationality tests under unstable environments, while \citet{odendahl2023evaluating} apply threshold models for state-dependent forecast evaluation.

The surge in inflation during and after the COVID-19 pandemic has sparked interest in inflation persistence and forecast accuracy. \citet{bianchi2023fiscal} develop a general-equilibrium model attributing inflation persistence to unfunded fiscal shocks during the pandemic. Similarly, \citet{koch2024we} identify factors behind inflation forecast errors during this period, such as stronger-than-expected demand recovery, labor market tightness, and fiscal stimulus.

Recent studies focus on central banks’ inflation forecasts, reflecting a global failure to predict inflation post-COVID-19. \citet{buturac2021measurement} provides a citation-based survey on measuring economic forecast accuracy, while \citet{binder2024central} survey recent literature on central bank forecast evaluations, focusing on major institutions like the Fed, ECB, Bank of England (BoE), and Bank of Canada (BoC).

Several studies relate inflation forecast errors to inflation targets. \citet{argiri2024evaluation} evaluate inflation forecasting by the ECB, BoE, and Fed, linking forecast bias to inflation targets. They regress forecasts on inflation targets and realized values, concluding that mid-term forecasts are primarily influenced by inflation targets. \citet{granziera2025bias} find that the ECB inflation forecasts exhibit state-dependent bias, with over-predictions when inflation is below the target and under-predictions when it is above. Unlike our study, their analysis estimates the inflation target threshold indirectly, as the ECB does not explicitly state its target level.

There are studies that also explore state-dependency of forecast bias of other economic indicators. \citet{sinclair2010can} investigate whether the Fed’s forecast efficiency is influenced by economic states, adding a dummy for NBER-dated recessions to the \cite{mincer1969evaluation} regression. Similarly, \citet{kpl2011} assess the state-dependency of Korean forecasts, using Statistics Korea’s ex-post recession indicators. \citet{xie2016time} determine economic states directly from data using a moving-average method, while \citet{granziera2025bias} apply \citet{odendahl2023evaluating}'s threshold estimation approach.

Few studies have evaluated the BoK forecasts. \citet{kpl2011} assess forecasts by Korean and international institutions, including the BoK, finding that GDP growth forecasts are overestimated during recessions, while inflation forecasts show no such bias.

%Cho and Kim (1999) \cite{ck1999} use half-yearly forecast data of several economic research institutes running from 1982 to 1997. They test whether the rational expectation hypothesis applies to those research institutes. Their research do not reject null hypothesis of Mincer-Zarnowitz test. Son and Kim (2008) \cite{sk2008} use forecasts made by various research institutes, including the Bank of Korea. They reveal the existence of information asymmetry between the BoK and other private research institutes. They also find an evidence that the BoK's forecast affects the expectations of private sector. 
%In addition, they estimate their own vector error correction model (VECM) and compare the out-of-sample performance with the existing forecasts. 
%Kim and Ahn (2015) \cite{ka2015} use monthly data of the survey firm \textit{Consensus Economics} since January 1995 until December 2013. They analyze whether the BoK's forecast has effects on variance of private sector's expectation. Until recently, the researches conclude that the BoK's inflation forecast is unbiased and efficient. However, to the best of our knowledge, there is no paper that thoroughly investigates the Bank of Korea's inflation forecast and relates its bias to inflation targeting.

\section{Data Description}

We construct half-yearly time series of forecasts and actual values for major macroeconomic variables in Korea, covering the period from H1 1999 to H2 2024. The dataset includes GDP growth, CPI inflation, and unemployment rates, sourced from various institutions. The BoK provides macroeconomic projections through its periodicals, \textit{Economic Outlook}, while Statistics Korea (KOSTAT) releases actual values of CPI and unemployment in corresponding survey reports. Lastly, realized values of the real GDP growth rate are announced by the BoK, and the forecasts are evaluated against these realized values. The construction of the vintage is detailed in Section 3.2.

\subsection{Forecast Data}

The BoK’s forecasts are accessed through its past press releases and \textit{Economic Outlook} periodicals.\footnote{Since there is no publicly available, readily usable dataset, we manually collected the BoK’s forecasts from these reports.} Table \ref{Announcement} summarizes the history of the BoK’s economic forecast announcements. The structure of the announcements, including the press release dates and the list of forecasted variables, has changed over time.

Before the first publication of \textit{Economic Outlook} in 2012, the BoK released its economic projections to the public through press releases, starting in April 1999. These announcements included forecasts of the real GDP growth rate, headline CPI inflation rate, percentage changes in the current account, and detailed components thereof.

For GDP, forecasts also covered components of expenditure, such as private consumption, private investment, and goods imports and exports. The inflation section included projections for both core inflation and headline CPI inflation. Labor market projections were added in December 2004, covering changes in the number of employed persons, the unemployment rate, and the employment-to-population ratio. Notably, the unemployment rate was based on the four-week job search definition established by the International Labour Organization. Lastly, the current account section forecasted the service account, goods account, and primary income account. All variables are reported as year-on-year (YoY) rates, with non-seasonally adjusted series unless explicitly stated otherwise.

Press releases occurred twice or three times a year—typically in April, July (when three announcements were made), and December. The forecast horizons varied by announcement month. For example, in December, forecasts extended up to five quarters ahead, while in July, forecasts generally extended only to the current year (a maximum of two quarters ahead).

A major revision in April 2009 fixed the forecast horizon to two years, though variations across release months persisted. To reduce market confusion, the press release timing was synchronized with the Monetary Policy Board’s (MPB) monetary policy decisions.

The publication of \textit{Economic Outlook} began in July 2012, with quarterly releases mirroring practices at other major central banks. Initially, reports were published in January, April, July, and October (2012–2019). Currently, they are released in February, May, August, and November, timed to follow MPB regular meetings.

\textit{Economic Outlook} provides projections for half-yearly and yearly macroeconomic variables, often including detailed subcomponents. For this study, we focus on semiannual forecasts of three major variables: the real GDP growth rate, CPI inflation rate, and unemployment rate. In particular, we pay special attention to inflation projections to examine the relationship between their bias and the inflation targeting regime, as the BoK sets the inflation target and produces the forecasts.

\begin{table}[t]
    \begin{adjustbox}{width=\columnwidth,center}
        \begin{tabular}{lllll}
        \hline\hline \\
            \textbf{Periods} &
              \begin{tabular}[c]{@{}c@{}}
                   \textbf{Press (publication)}  \\ \textbf{Month}
              \end{tabular} &
              \begin{tabular}[c]{@{}c@{}}
                   \textbf{Press (publication)}  \\ \textbf{Date}
              \end{tabular} &
              \textbf{Forecast Horizon} &
              \textbf{Forecast Variables} \\ \\ \hline \\
            1999-2001 &
              \begin{tabular}[c]{@{}l@{}}July (June), \\ December \\ (October)\end{tabular} &
              - &
              \begin{tabular}[c]{@{}l@{}}2 quarters - \\ 5 quarters\end{tabular} &
              \begin{tabular}[c]{@{}l@{}} Output growth, \\ prices, current account\end{tabular} \\ \\
            2002-2003 &
              \begin{tabular}[c]{@{}l@{}}April, July, \\ December\end{tabular} &
              - &
              \begin{tabular}[c]{@{}l@{}}2 quarters - \\ 5 quarters\end{tabular} &
              - \\ \\
            2004-2008 &
              July, December &
              - &
              \begin{tabular}[c]{@{}l@{}}2 quarters - \\ 5 quarters\end{tabular} &
              \begin{tabular}[c]{@{}l@{}}Employment section \\ is added \\ since December 2004\end{tabular} \\ \\
            Dec 2009-2012 &
              \begin{tabular}[c]{@{}l@{}}April, July, \\ December\end{tabular} &
              \begin{tabular}[c]{@{}l@{}}Following business day \\ after the *MPB \\ regular meeting\end{tabular} & \begin{tabular}{@{}l@{}}
                   Current and\\next years
              \end{tabular}  &
              - \\ \\
            Jul 2012-2019 &
              \begin{tabular}[c]{@{}l@{}}January, April, \\ July, October\end{tabular} &
              \begin{tabular}[c]{@{}l@{}}The same day \\ after the MPB \\ regular meeting\end{tabular} &
              \begin{tabular}{@{}l@{}}
                   Current and\\next years
              \end{tabular} &
              - \\ \\
            Nov 2019-present &
              \begin{tabular}[c]{@{}l@{}}February, May, \\ August, November\end{tabular} &
              - &
              \begin{tabular}{@{}l@{}}
                   Current and\\next years
              \end{tabular} &
              - \\ \\ \hline
        \end{tabular}
    \end{adjustbox}
    \caption{History of the BoK's forecast announcement. *MPB: Monetary Policy Board.}
    \label{Announcement}
\end{table}

\subsection{Realized Values}

In this subsection we clarify against which vintage the forecasts are evaluated. The BoK provides quarterly real GDP series, while Statistics Korea distributes monthly consumer price indices and labor market statistics (e.g., number of unemployed persons, labor force size). Since these data are available only as quarterly or annual rates, we convert their units into semiannual rates to match those of the forecasts and realizations.

An additional issue arises with the reported values. The BoK’s forecasts are rounded to the first decimal place, as are the actual quarterly and annual rates. For consistency, we also round the transformed half-yearly realized values to the first decimal place. The sources of the actual realized data are summarized in Table \ref{tbl:rvalue}.

\begin{table}[t]
	\centering
	\begin{adjustbox}{width=\columnwidth,center}
		\begin{tabular}{llll}
			\hline\hline
			\textbf{Variable} & \textbf{Frequency} & \textbf{Sample Period} & \textbf{Data Source}                         \\ \hline
			CPI               & Monthly            & Jan 1998-Jun 2024      & KOSTAT \textit{Consumer Price Index  }                \\
			Real GDP          & Quarterly          & Q1 1999-Q1 2024        & ECOS \textit{National Account 2.1.2.1.4.}             \\
			Unemployment Level & Monthly & Jan 2004-Jun 2024 & KOSTAT \textit{Economically Active Population Survey} \\
			Labor Force       & Monthly            & Jan 2004-Jun 2024      & KOSTAT \textit{Economically Active Population Survey} \\ \hline
		\end{tabular}
	\end{adjustbox}
	\caption{Actual realized values}
	\label{tbl:rvalue}
\end{table}

\paragraph{Real GDP}
Quarterly real GDP data is available from ECOS (Economic Statistics System) of the BoK. Specifically, the item \textit{2.1.2.1.4. GDP and GNI by Economic Activities (not seasonally adjusted, chained 2015 year prices, quarterly \& annual)} is used. The base year is 2020. While final estimates are primarily used, the values in H1 2023 and H2 2023 are preliminary estimates.

To derive biannual data, we sum the first and second quarters to calculate the first half-year real GDP. Similarly, summing the third and fourth quarters gives the second half-year real GDP. The growth rate is calculated on a YoY basis.

\begin{figure}[t]
	\centering
	\includegraphics[width=1\linewidth]{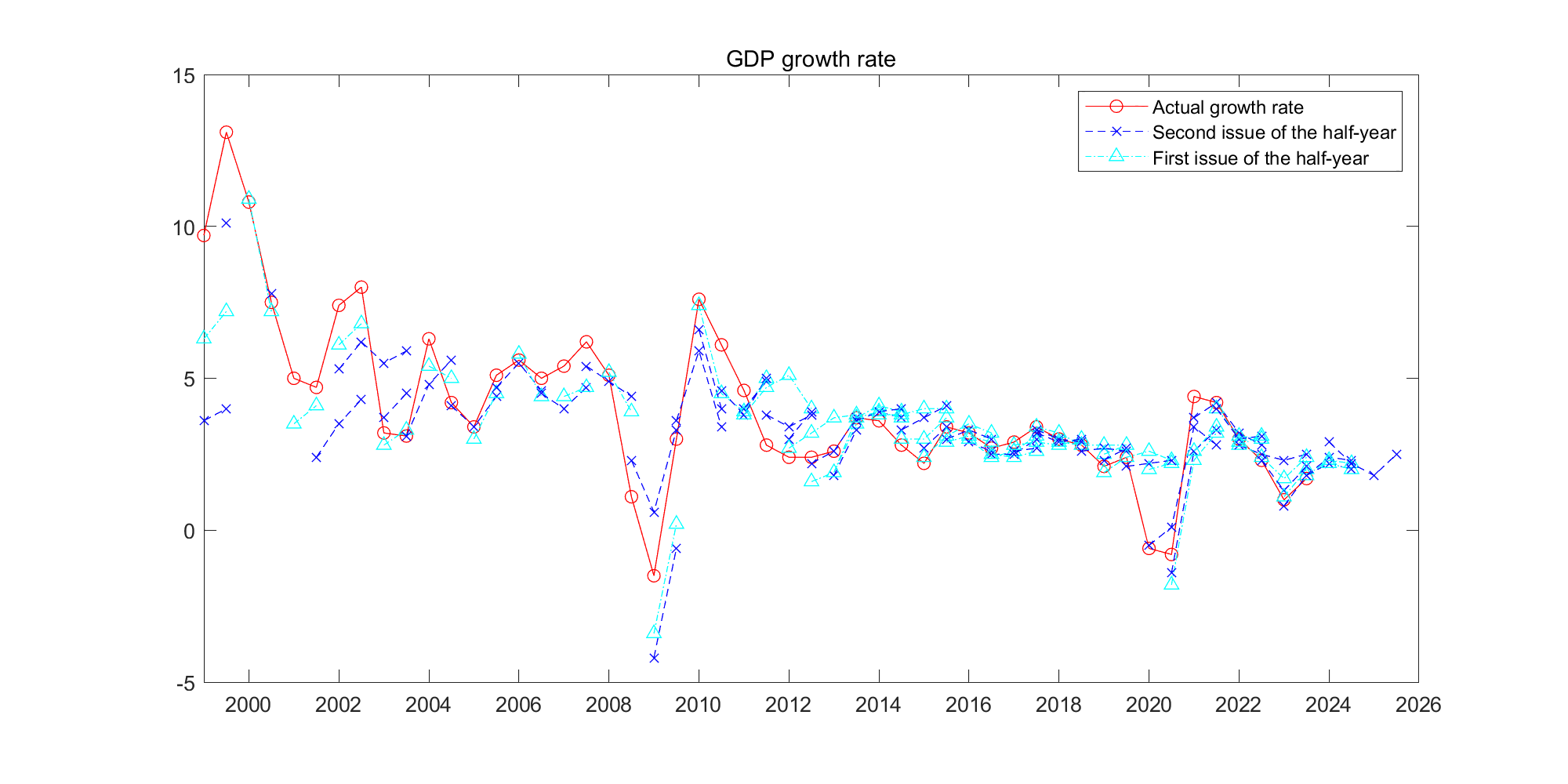}
	\caption{Forecasted and realized values of the real GDP growth rate. $\circ$: actual realized growth rates; $\triangle$: forecasts from the first EO issues; $\times$: forecasts from the second EO issues.}
	\label{gdp1}
\end{figure}

Figure \ref{gdp1} illustrates the forecast and realized values of the real GDP growth rate in Korea since 2012. The red line with circles represents the actual realized growth rate, while shorter dashed lines denote the forecast paths in each issue of the \textit{Economic Outlook}. Specifically, dashed lines with triangle markers indicate the first issues of the corresponding half-year (\textit{Economic Outlook - February} for the first half-year and \textit{Economic Outlook - July} for the second half-year). Dashed lines with x-markers represent forecasts from second issues.

The actual GDP growth rate fluctuates around 3\%, with notable dips. During the COVID-19 pandemic, it plummeted to a negative growth rate. The economy recovered from the recession in 2021. Recent BoK forecasts suggest that the growth rate is expected to stabilize around 2\% in the near future.

\paragraph{Unemployment Level and Labor Force.}
Statistics Korea provides up-to-date information on unemployment, employment, labor force, and related metrics on a monthly basis. The corresponding survey is called the \textit{Economically Active Population Survey} (EAPS). \textit{EAPS} reports quarterly and annual data at the end of every quarter and year, respectively. These variables are calculated as averages over three months for quarterly data or twelve months for annual data. Based on this, we derive the semiannual unemployment rate using the following formula:
\[unrate_{t}=\left(\frac{\sum_{m \in M_t} unemp_{m}}{\sum_{m \in M_t} labor_{m}}\right) \times 100\]
where $unrate$ is the unemployment rate, $unemp$ is the number of unemployed individuals, and $labor$ represents the labor force. $M_t$ denotes the set of months corresponding to the half-year $t$.

\begin{figure}[t]
	\centering
	\includegraphics[width=1\linewidth]{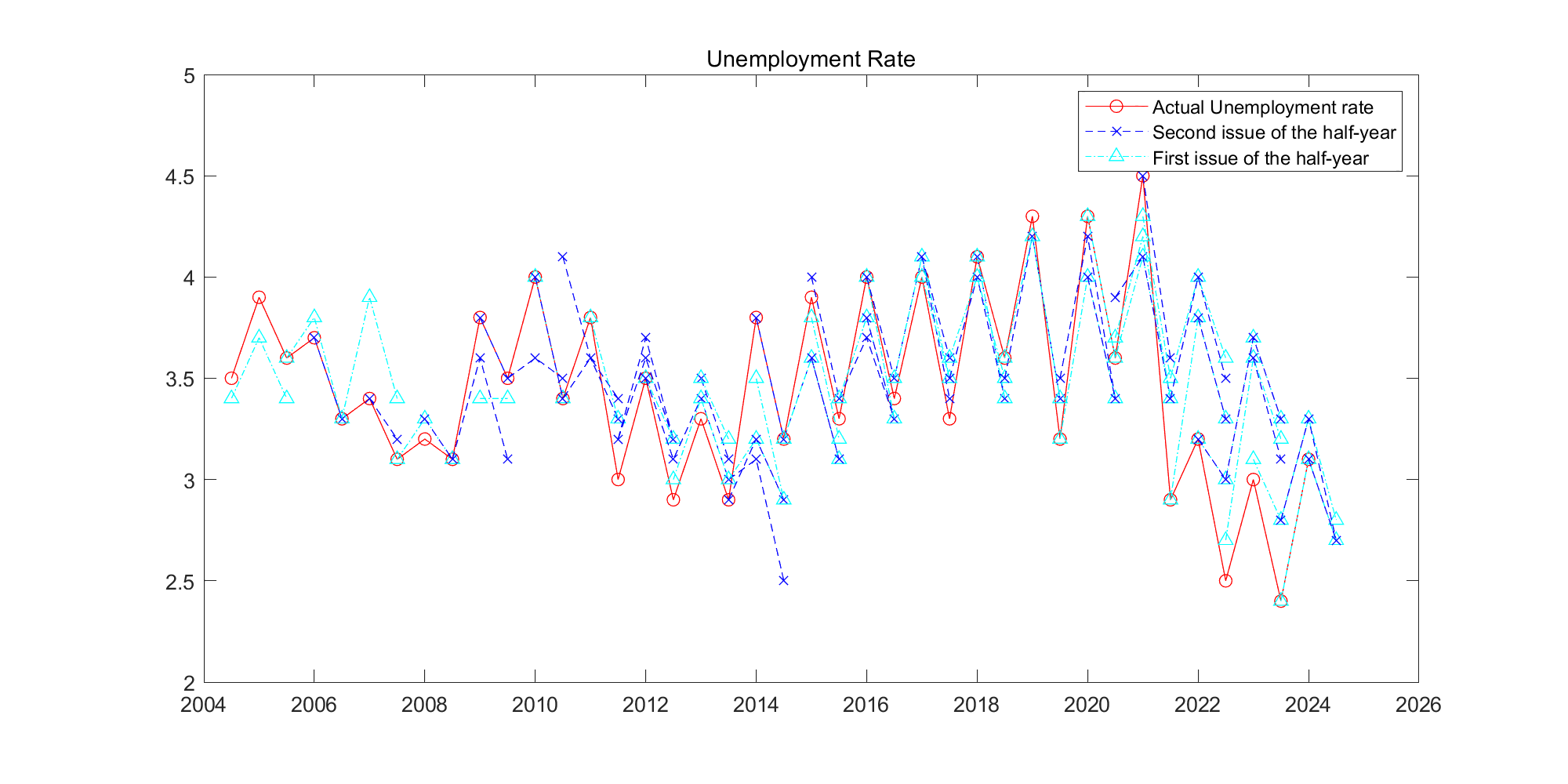}
	\caption{Forecasted and realized values of unemployment rate. $\circ$: actual realized unemployment rates; $\triangle$: forecasts from the first EO issues; $\times$: forecasts from the second EO issues.}
	\label{unrate1}
\end{figure}

Figure \ref{unrate1} illustrates the realized unemployment rates and their forecasts. A clear seasonality is evident in the unemployment rate, which tends to be higher in the first half-year compared to the second. The forecasts reflect this seasonality. The unemployment rate averaged around 3.5\% until 2021. However, it has recently decreased, leading to significant forecast errors.

\paragraph{Consumer Price Index.}
Monthly CPI is announced each following month by KOSTAT through the \textit{Consumer Price Index} press release. Historical indices, with base year of 2020, are publicly accessible via KOSIS (Korean Statistical Information Service). According to KOSTAT, annual CPI inflation is calculated as the ratio of the average monthly indices of the current year to those of the previous year. Using this calculation, we define the half-yearly inflation rates as follows:
\[
    \pi^{CPI}_{t}=\left(\frac{\sum_{m \in M_t} CPI_{m}}{\sum_{m \in M_{t-2}} CPI_{m}}-1\right) \times 100
\] 
Here, $M_t$ denotes the set of months corresponding to the half-year $t$.

\begin{figure}[t]
	\centering
	\includegraphics[width=1\linewidth]{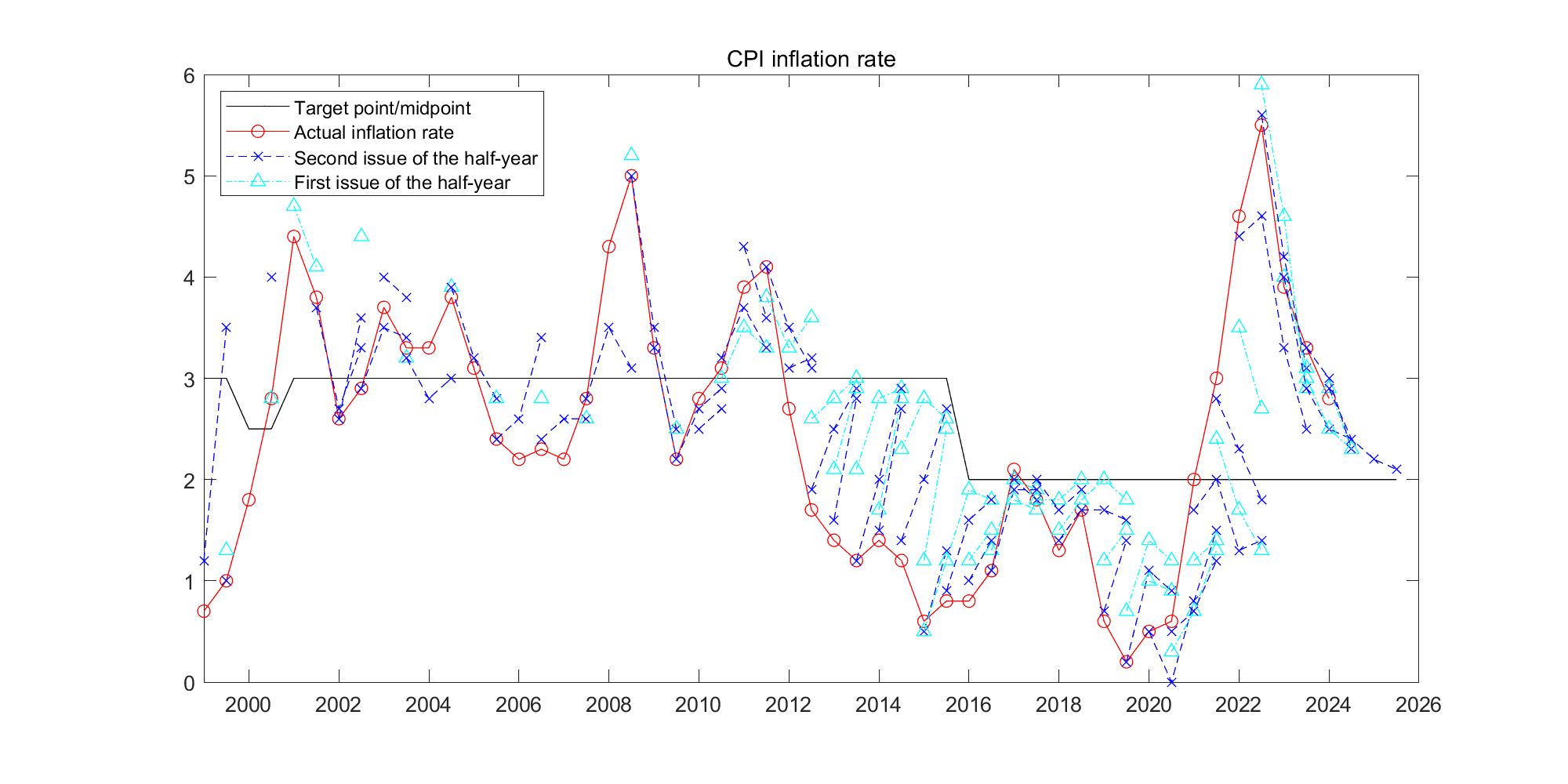}
	\caption{Forecasted and realized values of inflation rate. The black solid line indicates the target point or the midpoint of target range. $\circ$: actual realized inflation rates; $\triangle$: forecasts from the first EO issues; $\times$: forecasts from the second EO issues.}
	\label{infl1}
\end{figure}

Figure \ref{infl1} illustrates the forecasts and realized values of the CPI inflation rate in Korea since 2012. The black solid line represents the midpoint of the Bank of Korea’s inflation target range or its target point, while the dashed lines depict projection paths from each respective issue. The forecast horizon varies, ranging from a minimum of 2 to a maximum of 4 half-year periods ahead.

Several patterns emerge from the figure. The initial points (short-term forecasts) of each issue tend to be relatively close to the realized values. However, as the forecast horizon extends further, the forecast line converges toward a specific point. Notably, the BoK’s inflation target was 3\% until 2015 and has been 2\% since then.

Figure \ref{infl1} also reveals a systematic bias in the BoK’s forecasts. When actual inflation is below the target, the inflation forecast tends to be overestimated, and vice versa.

\subsection{Timeline of Forecast and Realization}

The structure of our dataset is complicated due to the asynchrony between the frequency of the forecasted variable and the forecast horizon. For illustration, let the realized CPI inflation rate in period $t$ be denoted by $y_t$, and let $y_{h,t}$ represent the $h$-quarters-ahead prediction for $y_t$. It is important to note that $h$ is on a quarterly basis, whereas the frequency of the time series $t$ is half-yearly. Specifically, if a forecast is made less than three months before the end of the half-year (i.e., June or December), we index the forecast by $h=0$ and classify $y_{0,t}$ as a zero-quarter-ahead forecast of $y_t$ (i.e., a nowcast).

For example, let $t = \text{H1 2021}$ (the first half of 2021) and $h = 0$. Then, $y_{0,\text{H1 2021}}$ denotes the predicted value released to the public in Q2 2021 (\textit{Economic Outlook – May 2021}). Since the time lag between May 2021 and July 1, 2021 is less than three months, this forecast is classified as $h = 0$. Similarly, the time lag between the publication of $y_{1,\text{H1 2021}}$ and the realization of $y_{\text{H1 2021}}$ is approximately five months, so it is classified as a one-quarter-ahead forecast, and so on.

\begin{figure}[t]
	\centering
\begin{tikzpicture}[scale=0.85]
    \draw[->] (-0.5,0) -- (14,0) node[anchor=north west] {};
    %\draw[-] (-2,1) -- (14,1) node[anchor=north west] {};

    \foreach \x in {0,6,12}
        \draw (\x,10pt) -- (\x,0pt) node[anchor=north] {};
    \foreach \x in {0,6,12}
        \draw (\x,0pt) -- (\x,-10pt) node[anchor=north]{};
    \foreach \x in {2,4,8,10}
        \draw (\x,5pt) -- (\x,0pt) node[anchor=north] {};

    %\node[above] at (-1,0) {$EO^{second}_{t-1}$};
    %    \node[above] at (-1,1) {$y_{0,t-1}$};
    %    \node[above] at (-1,1.5) {$y_{2,t}$};
    %    \node[above] at (-1,2) {$y_{4,t+1}$};
    %    \node[above] at (-1,3) {};

    \node[above] at (2,2.2) {$EO^{1st}_{t}$};
        \node[above] at (2,0.3) {$\begin{pmatrix}y_{1,t}\\y_{3,t+1}\\y_{5,t+2}\end{pmatrix}$};
        %\node[above] at (2,1.5) {$y_{3,t+1}$};
        %\node[above] at (2,2) {$y_{5,t+2}$};
        \node[above] at (2,4) {};
    \node[above] at (4,2.2) {$EO^{2nd}_{t}$};
        \node[above] at (4,0.3) {$\begin{pmatrix}y_{0,t}\\y_{2,t+1}\\y_{4,t+2}\end{pmatrix}$};
        %\node[above] at (4,1.5) {$y_{2,t+1}$};
        %\node[above] at (4,2) {$y_{4,t+2}$};
        \node[above] at (4,4) {};
    \node[above] at (8,2.2) {$EO^{1st}_{t+1}$};
        \node[above] at (8,0.3) {$\begin{pmatrix}y_{1,t+1}\\y_{3,t+2}\\y_{5,t+3}\end{pmatrix}$};
        %\node[above] at (8,1.5) {$y_{3,t+2}$};
        %\node[above] at (8,2) {$y_{5,t+3}$};
        \node[above] at (8,4) {};
    \node[above] at (10,2.2) {$EO^{2nd}_{t+1}$};
        \node[above] at (10,0.3) {$\begin{pmatrix}y_{0,t+1}\\y_{2,t+2}\\y_{4,t+3}\end{pmatrix}$};
        %\node[above] at (10,1.5) {$y_{2,t+2}$};
        %\node[above] at (10,2) {$y_{4,t+3}$};
        \node[above] at (10,4) {};

    \node[above] at (14,2.2) {$EO^{1st}_{t+2}$};
        \node[above] at (14,0.3) {$\begin{pmatrix}y_{1,t+2}\\y_{3,t+3}\\y_{5,t+4}\end{pmatrix}$};
        %\node[above] at (13,1.5) {$y_{3,t+3}$};
        %\node[above] at (13,2) {$y_{5,t+4}$};
        \node[above] at (14,4) {};
        
    \node[below] at (9,-0.5) {};
    \node[below] at (3,-0.5) {};
    
    \node[below] at (0,-0.3) {$t-1$};
        \node[below] at (0.2,-1) {$y_{t-1}$};
        \node[below] at (0.2,-1.5) {$e_{h,t-1}$};
    \node[below] at (6,-0.3) {$t$};
        \node[below] at (6,-1) {$y_t$};
        \node[below] at (6,-1.5) {$e_{h,t}$};
    \node[below] at (12,-0.3) {$t+1$};
        \node[below] at (12,-1) {$y_{t+1}$};
        \node[below] at (12,-1.5) {$e_{h,t+1}$};
        \node[below] at (12,-2) {};

    \node[above] at (-2,1.5) {Forecast};
    \node[below] at (-2,-1) {Realization};
\end{tikzpicture}
\caption{Timeline of Forecasts and Realizations of Macroeconomic Variables}
\label{fig:diag}
\end{figure}
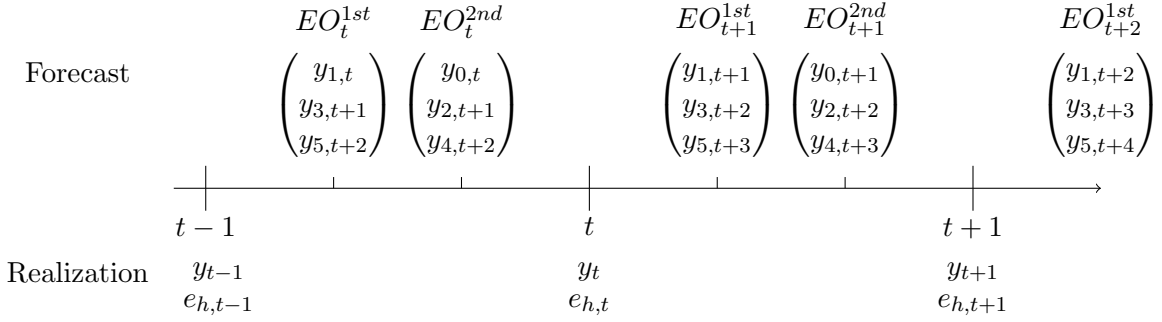

To understand the structure of our data, it is essential to clarify the timeline of forecasts and realizations, which we summarize in Figure \ref{fig:diag}. The horizontal arrow depicts the time axis. Variables above the axis correspond to forecasts from the respective \textit{Economic Outlook} issues, while those below the axis indicate the corresponding realizations. Actual values are realized and publicly released each period. The first issue of the \textit{Economic Outlook} for period $t$, denoted as $EO_t^{1st}$, and the second issue, $EO_t^{2nd}$, are both published during $t$. To highlight their chronological order, these issues are placed in separate bins along the timeline.

The $EO_t^{1st}$ report provides forecasts for $y_t$, $y_{t+1}$, and $y_{t+2}$, denoted by $y_{1,t}$, $y_{3,t+1}$, and $y_{5,t+2}$, respectively. The first subscript in each notation indicates the forecast horizon: for example, $y_{1,t}$ is a one-quarter-ahead forecast of $y_t$, while $y_{3,t+1}$ is a three-quarter-ahead forecast of $y_{t+1}$. Similarly, $EO_t^{2nd}$ provides updated forecasts for $y_t$, $y_{t+1}$, and $y_{t+2}$, but with shorter horizons: $y_{0,t}$, $y_{2,t+1}$, and $y_{4,t+2}$.

Forecast errors for all horizons become observable simultaneously with the announcement of the actual value. For instance, at time $t$, forecast errors $e_{h,t}$ for all horizons $h = 0, 1, \dots, 5$ are realized once the actual value $y_t$ is released.

The publication date of the \textit{Economic Outlook} has varied over time (Table \ref{Announcement}). Currently, since 2019, the BoK is releasing its forecasts in February, May, August, and November.\footnote{Once the time index is fixed, the number of forecasts available for each period is determined. The maximum horizon for the CPI inflation rate is 6 quarters (Figure \ref{infl1}).} We classify February and August issues as the first \textit{Economic Outlook} (EO) issue of the corresponding half year; May and November are classified as the second issues.

However, prior to 2012, forecasts were published only two or three times per year, resulting in missing values. For example, see the first row of Table \ref{Announcement}. From 1999 to 2001, reports were issued in July and December, which are classified as $EO_t^{2nd}$. Since $EO_t^{1st}$ was not published during these years, the values of $y_{1,t}$, $y_{3,t+1}$, $\dots$, $y_{5,t+3}$ are missing.

%The authorities do not report the relevant indices in half-yearly variables, so we properly aggregate and transform monthly or quarterly data to match the units of the forecast and the realization. As a result, the announcements in July and January are significant since we can calculate half-yearly variables at that time.

\subsection{Inflation Targeting}
\begin{figure}[ht]
	\centering
	\includegraphics[width=1\linewidth]{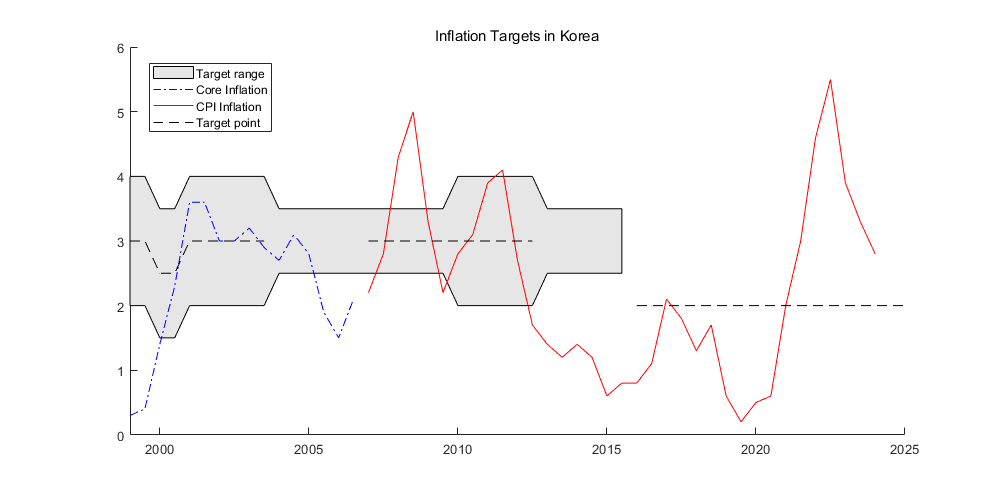}
	\caption{Inflation Targeting and Price Indices}
	\label{fig:InflationTarget}
\end{figure}

Figure \ref{infl1} shows that the BoK’s inflation forecasts appear biased and may be related to its inflation target. Here, we briefly outline the history of inflation targeting in Korea, which is summarized in Figure \ref{fig:InflationTarget}.

At the end of 1997, Korea transitioned its monetary policy framework from monetary targeting to inflation targeting. During the first two years of inflation targeting, the targets were based on the Headline CPI. In 1998, the inflation target was set at $9\pm1\%$, and it was adjusted downward to $3\pm1\%$ in 1999.

From 2000 to 2006, the targets were set based on core inflation, but CPI inflation became the standard again in 2007. Between 2000 and 2003, the targets were set annually: in 2000, the target was $2.5\pm1\%$, and from 2001 to 2003, it was $3\pm1\%$.

In 2004, considering the transmission lag of monetary policy, the BoK adopted a mid-term (three-year) inflation targeting horizon. From 2004 to 2006, the target range was set as the annual average of $2.5\%$–$3.5\%$ for the core inflation rate. From 2007 to 2009, it shifted to $3\pm0.5\%$ for CPI inflation.

During the period 2010–2012, the target band was expanded to $3\pm1\%$ to reflect global market uncertainty caused by the 2009 financial crisis. Subsequently, it was adjusted to $2.5\%$–$3.5\%$ for 2013–2015. Finally, since 2016, the BoK has maintained a fixed inflation target of $2\%$.

\section{Forecast Evaluation}

We define the forecast errors as:
\[e_{h,t}=y_t-y_{h,t},\]
where $y_{h,t}$ is the BoK’s $h$-quarters-ahead forecast for a macroeconomic variable $y_t$ of interest in period $t$. In the literature, $y_t$ is also referred to as the \textit{target} of the forecast. However, to avoid confusion with inflation targeting, we do not adopt this convention.

%In the  we also define forecast revisions. Forecast revisions can be used to evaluate forecast optimality. We present those related test results in the Appendix, and focus on the forecast errors in this section.

\begin{figure}[t]
    \centering
    \includegraphics[width=1\linewidth]{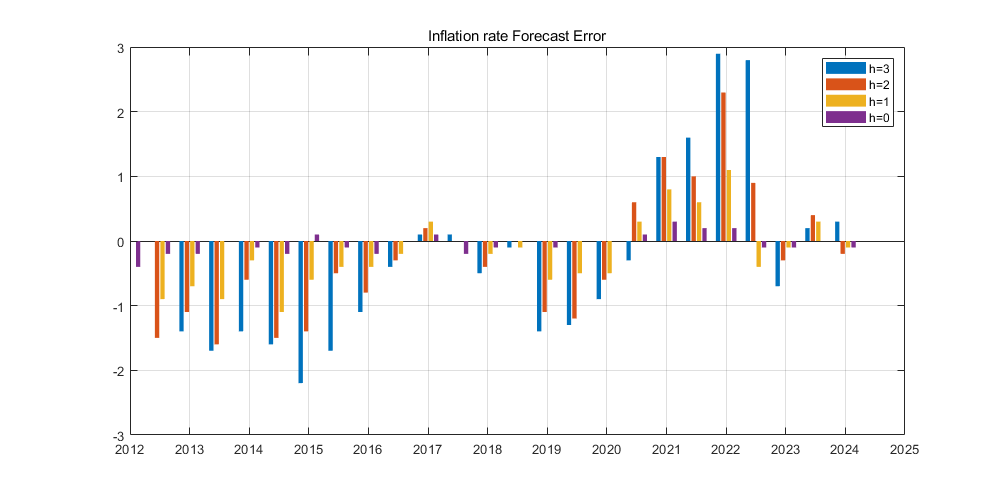}
    \caption{CPI Inflation Rate Forecast Error. The x-axis represents half-year periods. For each time period, there are four bars corresponding to forecast errors for each forecast horizon. The leftmost bar represents the forecast error for $h=3$, followed sequentially by forecast errors for shorter horizons.}
    \label{fe_infl}
\end{figure}

Figure \ref{fe_infl} illustrates the inflation forecast errors. The y-axis represents the percentage point difference between the forecast and the realization, while the x-axis corresponds to the forecast target’s period, with a frequency of 6 months. Each period contains four bars, representing $h = 3$, $h = 2$, $h = 1$, and $h = 0$ from left to right. The figure reveals notable patterns. First, forecast errors exhibit high persistence, maintaining their sign over extended periods when fixing the forecast horizon. Second, there is a positive correlation in forecast errors across horizons, indicating that errors at different horizons tend to move in the same direction. However, the magnitude of forecast errors generally decreases as the forecast horizon shortens, with errors approaching zero as $h$ approaches zero. Lastly, by comparing Figure \ref{fe_infl} with Figure \ref{fig:InflationTarget}, we find that when inflation is below the target, forecast errors tend to be negative (i.e., overestimating and biased toward the target) during 2013--2016 and 2019--2020. Conversely, when inflation is above the target in 2021--2022, forecast error tend to be positive (i.e., underestimating and again biased toward the target). The forecast errors for GDP growth and unemployment rates are presented in Figure \ref{fig:forecast_error} in the Appendix.

Following the literature, we define forecast bias as the unconditional expectation of the forecast error. A forecast systematically overestimates or underestimates the realization if the sign of the bias, $E[e_{h,t}] \equiv E[y_t - y_{h,t}]$, is negative (i.e., the forecast is, on average, higher than the realization) or positive, respectively.
To evaluate the unbiasedness of the BoK’s forecasts, we test whether the forecast errors satisfy the mean-zero property. Let $y_t - y_{h,t} \equiv e_{h,t} = \delta_h + \nu_{h,t}$. Then we test:
\begin{equation*}
H_0: \delta_h = 0.
\end{equation*}
for each $h=0,1,2,\dots,4$.

We rearrange the data so that each forecast time series shares a common set of forecast horizons. The tests are conducted separately for each horizon. Pooling the data across horizons and structuring the sample as a panel is not straightforward, as the forecast errors may exhibit cross-sectional correlation across horizons. This implies that a forecaster could have made similar errors when revising projections for the same target period. Such dependencies can affect the validity of hypothesis testing. \citet{clements2007evaluation} propose a robust test statistic to account for cross-horizon dependencies when testing $H_0: \delta_h = 0$.

There is a caveat regarding the mean-zero forecast error test. Even if a forecast systematically underestimates or overestimates the variable of interest under certain economic conditions, the test may fail to reject the null hypothesis if positive and negative errors offset each other.

\begin{figure}[ht]
    \centering
    \includegraphics[width=1\linewidth]{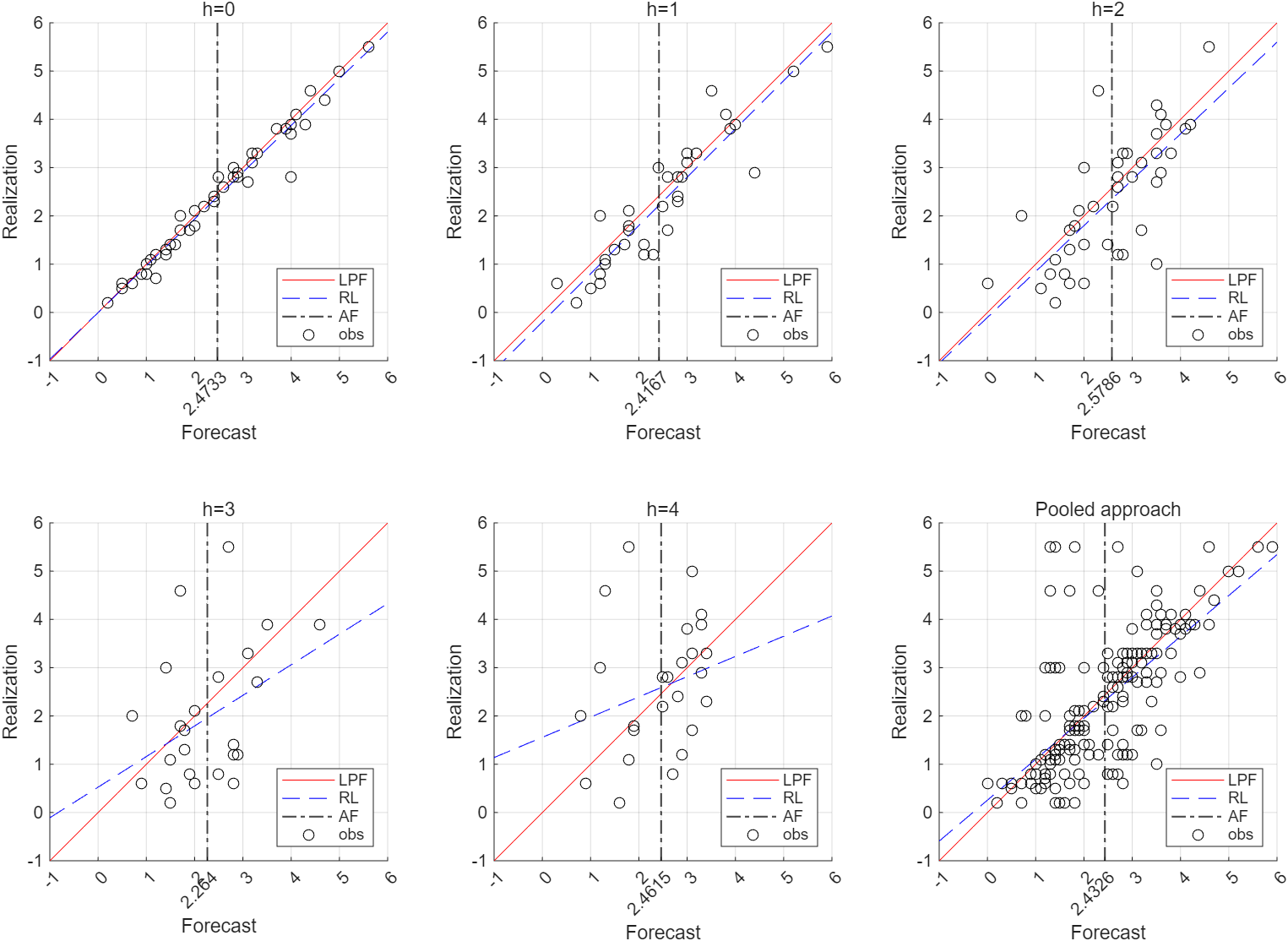}
    \caption{Conditional Mean of CPI Inflation Rate Forecasts by Horizon and Pooled Approach. The four-digit decimal numbers represent the sample averages of the forecasts (AF). If the regression line (RL) crosses the Line of Perfect Forecasting (LPF) at these values, this indicates that the average forecast error is zero.}
    \label{fig:Mincer}
\end{figure}

One of the seminal works in the forecast evaluation literature is the handbook chapter by \citet{mincer1969evaluation}. Following their approach, we display the ordered pairs $(y_{h,t}, y_t)$ of forecasts and realizations in Figure \ref{fig:Mincer}, with forecasts on the x-axis and realizations on the y-axis. Each panel corresponds to a different forecast horizon. The black circles represent individual forecast–realization pairs. The red line denotes the $45^\circ$ line through the origin, referred to as the \textit{Line of Perfect Forecast} (LPF). The blue dashed line indicates the estimated regression line (RL) from the \citet{mincer1969evaluation} equation: $y_t = \alpha_h + \beta_h y_{h,t} + v_{h,t}$. The decimal number labeled on the x-axis of each panel indicates the average forecast (AF). If the regression line (RL) does not intersect the Line of Perfect Forecast (LPF) at this point, the forecast is considered biased.

\citet{mincer1969evaluation} also propose an alternative test of unbiasedness based on the estimated regression line. They argue that a \textit{good} forecast should lie close to the Line of Perfect Forecast (LPF) not only at the average forecast value, but across the entire range of forecast values. For example, in the $h = 4$ panel of Figure \ref{fig:Mincer}, the regression line (RL) intersects the LPF near the average forecast value of 2.4615, suggesting little evidence of bias in the $h = 4$ forecasts. However, for forecast values below 2.4615, the RL lies above the LPF, and for values above 2.4615, it lies below, indicating potential systematic deviations. To formally assess this, \citet{mincer1969evaluation} propose testing the joint null hypothesis for the regression equation:\[H_0: \alpha_h = 0, \beta_h = 1\]
for each $h = 0, 1, 2, \dots, 4$. Under this null hypothesis, the regression line lies exactly on the LPF.

\citet{holden1990testing} argue that the null hypothesis of the Mincer-Zarnowitz test is a sufficient condition for unbiasedness. They show, instead, that the necessary and sufficient condition for unbiasedness is $H_0: \delta_h = 0$ where $e_{h,t} \equiv y_t - y_{h,t} = \delta_h + \nu_{h,t}$. It is noteworthy that the Holden-Peel test is implied by the Mincer-Zarnowitz test. In particular, the Holden-Peel test corresponds to the Mincer-Zarnowitz test under the restriction $\beta = 1$.

In Figure \ref{fig:Mincer}, we observe that forecasts for shorter horizons, such as $h = 0$ and $h = 1$, are densely distributed along the 45-degree line, indicating closer alignment with realizations. However, for longer forecast horizons ($h = 3, 4$), the forecasts exhibit less variation, and the regression line deviates from the 45-degree line.

%Forecast is unbiased if for all possible realization $y=x \in X \subset \mathbb{R}$, $$Bias(y_{h}|y=x)=E[y_h-y|y=x]=0.$$ Hence, assuming linear CEF $E[y_h|y=x]=\alpha+\beta x$, the unbiasedness of $y_h$, for each $h=0,1,2,...$, holds if and only if $\alpha =0$ and $\beta=1$. Thus our test is the following:
%\[
%y_{h,t}=\alpha_h+\beta_h y_t+e_{h,t}
%\]\[
%H_0:\alpha_h=0,\beta_h=1
%\]
%for each $h=0,1,2,...$

\subsection{Test Results}

We now present the results of the unbiasedness tests for the BoK’s forecasts. The \citet{newey1994automatic} HAC estimator is used to calculate the test statistics. Table \ref{tbl:MZtest} displays the test results for inflation forecasts and other variables. The table is divided into two blocks: the left block contains the results of the Mincer and Zarnowitz test, while the right block contains the results of the Holden-Peel t-test. Tests are conducted for each horizon, $h = 0, 1, 2, 3, 4$, and the abbreviation “\textbf{N}” indicates the number of observations.

Although forecasts are available for horizons up to $h = 6$, the number of observations for these horizons is limited. Therefore, they are excluded from the table. In each block, the first column shows the estimated test statistics, and the second column provides the p-values of these statistics.

At a test size of 0.05, we fail to reject $H_0: \alpha_h = 0, \beta_h = 1$ for all horizons except $h = 4$. This result appears counter-intuitive. Similarly, at a test size of 0.05, we fail to reject $H_0: \delta_h = 0$ for all horizons, whether using one-sided or two-sided t-tests.

\begin{table}[t]
	\begin{adjustbox}{width=1\columnwidth,center}
		\begin{tabular}{lrlrlccc|rlcc}
			\multicolumn{12}{c}{Mincer-Zarnowitz and Holden-Peel test} \\ \hline\hline
			\multicolumn{2}{c}{}&\multicolumn{6}{c|}{$y_t=\alpha_h+\beta_h y_{h,t}+u_{h,t}$} & \multicolumn{4}{c}{$e_{h,t}=\delta_h+v_{h,t}$}
			\\ 
			\multicolumn{1}{l}{\textbf{Forecast}}& & \multicolumn{6}{c|}{\multirow{2}{*}{$H_0:\alpha_h=0,\beta_h=1$}} & \multicolumn{4}{c}{\multirow{2}{*}{$H_0:\delta_h=0$}} \\
			\multicolumn{1}{l}{\textbf{Horizon}}&\multicolumn{7}{c|}{} & \multicolumn{4}{c}{}   \\ \hline
			\multirow{2}{*}{CPI} & \multicolumn{2}{c}{\multirow{2}{*}{$\alpha_h$}} & \multicolumn{2}{c}{\multirow{2}{*}{$\beta_h$}} & \multicolumn{1}{c}{\multirow{2}{*}{\textbf{Wald}}} & \multicolumn{1}{c}{\multirow{2}{*}{\textbf{p-value}}} & \multicolumn{1}{c|}{\multirow{2}{*}{\textbf{N}}} & \multicolumn{2}{c}{\multirow{2}{*}{$\delta_h$}} & \multicolumn{1}{c}{\multirow{2}{*}{\textbf{t}}} & \multicolumn{1}{c}{\multirow{2}{*}{\textbf{p-value}}} \\
			& \multicolumn{2}{c}{}                       & \multicolumn{2}{c}{}                      & \multicolumn{1}{c}{}                               & \multicolumn{1}{c}{}                                  & \multicolumn{1}{c|}{}                            & \multicolumn{2}{c}{}                       & \multicolumn{1}{c}{}                            & \multicolumn{1}{c}{}                                  \\
			0                 & 0.00                & (0.03)               & 0.97               & (0.02)               & 3.97                                               & 0.14                                                  & 45                                              & -0.08               & (0.04)               & -1.86                                           & 0.06                                                  \\
			1                 & -0.20               & (0.18)               & 1.00               & (0.05)               & 3.13                                               & 0.21                                                  & 36                                              & -0.20               & (0.12)               & -1.65                                           & 0.10                                                  \\
			2                 & -0.10               & (0.31)               & 0.95               & (0.08)               & 2.13                                               & 0.34                                                  & 42                                              & -0.23               & (0.21)               & -1.10                                           & 0.27                                                  \\
			3                 & 0.52                & (0.47)               & 0.63               & (0.18)               & 4.01                                               & 0.13                                                  & 25                                              & -0.30               & (0.47)               & -0.64                                           & 0.52                                                  \\
			4                 & 1.56                & (0.71)               & 0.42               & (0.19)               & 11.63                                              & \textbf{0.00}                                         & 26                                              & 0.13                & (0.42)               & 0.30                                            & 0.76                                                  \\ 
			\multicolumn{8}{l|}{} & \multicolumn{4}{c}{}   \\ 
			\multirow{2}{*}{GDP} & \multicolumn{2}{c}{\multirow{2}{*}{$\alpha_h$}} & \multicolumn{2}{c}{\multirow{2}{*}{$\beta_h$}} & \multicolumn{1}{c}{\multirow{2}{*}{\textbf{Wald}}} & \multicolumn{1}{c}{\multirow{2}{*}{\textbf{p-value}}} & \multicolumn{1}{c|}{\multirow{2}{*}{\textbf{N}}} & \multicolumn{2}{c}{\multirow{2}{*}{$\delta_h$}} & \multicolumn{1}{c}{\multirow{2}{*}{\textbf{t}}} & \multicolumn{1}{c}{\multirow{2}{*}{\textbf{p-value}}} \\
			& \multicolumn{2}{c}{}                       & \multicolumn{2}{c}{}                      & \multicolumn{1}{c}{}                               & \multicolumn{1}{c}{}                                  & \multicolumn{1}{c|}{}                            & \multicolumn{2}{c}{}                       & \multicolumn{1}{c}{}                            & \multicolumn{1}{c}{}                                  \\
			-1                & 0.53                & (0.33)               & 0.97               & (0.08)               & 6.50                                               & \textbf{0.04}                                         & 27                                              & 0.43                & (0.17)               & 2.55                                            & \textbf{0.01}                                         \\
			0                 & 0.32                & (0.31)               & 1.07               & (0.10)               & 4.70                                               & 0.10                                                  & 44                                              & 0.54                & (0.28)               & 1.91                                            & 0.06                                                  \\
			1                 & -0.23               & (0.60)               & 1.14               & (0.22)               & 1.60                                               & 0.45                                                  & 35                                              & 0.23                & (0.22)               & 1.03                                            & 0.30                                                  \\
			2                 & 0.01                & (0.56)               & 1.08               & (0.16)               & 0.80                                               & 0.67                                                  & 41                                              & 0.28                & (0.35)               & 0.80                                            & 0.42                                                  \\
			3                 & 0.46                & (0.97)               & 0.69               & (0.29)               & 49.30                                              & \textbf{0.00}                                         & 24                                              & -0.50               & (0.13)               & -3.91                                           & \textbf{0.00}                                         \\
			4                 & 0.23                & (1.16)               & 0.85               & (0.29)               & 5.57                                               & 0.06                                                  & 25                                              & -0.33               & (0.16)               & -2.01                                           & \textbf{0.04}                                         \\ 
			\multicolumn{8}{l|}{} & \multicolumn{4}{c}{}   \\ 
			\multirow{2}{*}{Unemployment} & \multicolumn{2}{c}{\multirow{2}{*}{$\alpha_h$}} & \multicolumn{2}{c}{\multirow{2}{*}{$\beta_h$}} & \multicolumn{1}{c}{\multirow{2}{*}{\textbf{Wald}}} & \multicolumn{1}{c}{\multirow{2}{*}{\textbf{p-value}}} & \multicolumn{1}{c|}{\multirow{2}{*}{\textbf{N}}} & \multicolumn{2}{c}{\multirow{2}{*}{$\delta_h$}} & \multicolumn{1}{c}{\multirow{2}{*}{\textbf{t}}} & \multicolumn{1}{c}{\multirow{2}{*}{\textbf{p-value}}} \\
			& \multicolumn{2}{c}{}                       & \multicolumn{2}{c}{}                      & \multicolumn{1}{c}{}                               & \multicolumn{1}{c}{}                                  & \multicolumn{1}{c|}{}                            & \multicolumn{2}{c}{}                       & \multicolumn{1}{c}{}                            & \multicolumn{1}{c}{}                                  \\
			0                 & -0.04               & (0.22)               & 1.01               & (0.07)               & 2.27                                               & 0.32                                                  & 36                                              & -0.03               & (0.02)               & -1.43                                           & 0.15                                                  \\
			1                 & -0.48               & (0.28)               & 1.11               & (0.06)               & 3.27                                               & 0.19                                                  & 30                                              & -0.10               & (0.07)               & -1.33                                           & 0.18                                                  \\
			2                 & -0.51               & (0.49)               & 1.13               & (0.12)               & 1.79                                               & 0.41                                                  & 35                                              & -0.04               & (0.07)               & -0.55                                           & 0.58                                                  \\
			3                 & 0.04                & (0.81)               & 0.97               & (0.21)               & 0.63                                               & 0.73                                                  & 25                                              & -0.07               & (0.13)               & -0.54                                           & 0.59                                                  \\
			4                 & 0.81                & (0.73)               & 0.72               & (0.21)               & 7.60                                               & \textbf{0.02}                                         & 23                                              & -0.15               & (0.11)               & -1.37                                           & 0.17                                                 \\ \hline
		\end{tabular}
	\end{adjustbox}
\caption{Mincer and Zarnowitz (1969) and Holden and Peel (1990) tests for CPI Inflation, GDP Growth, and Unemployment Rate Forecasts.}
	\label{tbl:MZtest}
\end{table}

Table \ref{tbl:MZtest} also provides the same analysis for GDP growth and unemployment rates. We find evidence of bias in forecasts with the shortest horizon, $h = -1$, and with longer horizons, $h = 3, 4$, rejecting both $H_0: \alpha_h = 0, \beta_h = 1$ and $H_0: \delta_h = 0$. The BoK under-predicts the growth rate when forecasting one quarter ahead ($h = -1$) and over-predicts the target variables when forecasting three or four quarters earlier.\footnote{GDP forecasts may have $h = -1$ because it takes at least 28 days to estimate the current quarter’s GDP. For example, if the \textit{Economic Outlook - January 2018} forecasts H2 2017’s GDP growth rate and the advance quarterly estimate for Q4 2017 has not yet been announced, such a forecast would have a horizon of $h = -1$.} For horizons $h = 0, 1, 2$, we fail to reject the null hypotheses at a test size of 0.05. Finally, there is little evidence of bias in unemployment forecasts.

\subsection{State Dependent Analysis}

The Bank of Korea Act explicitly states that the objective of monetary policy is price stability. For this reason, we focus on inflation rate forecasts rather than on other macroeconomic variables such as the growth rate or the unemployment rate. Since the inflation rate is itself a policy target, inflation forecasts may be influenced by policy considerations, potentially making it difficult to maintain objectivity.

\citet{granziera2025bias} argue that the ECB's inflation forecasts over-predict inflation at intermediate forecast horizons when the inflation rate is below the target. In their analysis, the state of the economy is defined by whether the inflation rate is above or below the inflation target at the time of the forecast. They estimate state-dependent bias using a dummy variable generated from this definition.

Similarly, we define the state of the economy as follows:
\[d_{h,t} = I\{\pi_{h,t} \leq \pi_{h,t}^{target}\},\]
where $d_{h,t} = 1$ if the realized inflation rate $\pi_{h,t}$ is below the inflation target $\pi_{h,t}^{target}$ at the time of the forecast. For example, if $t=\text{H1 2021}$ and $h=0$, that is the BoK \textit{nowcasts} H1 2021 inflation rate, then $\pi_{h,t}^{target}$ is the actual inflation rate in Q2 2021. If $h=1$, then Q1 2021, and so on. The inflation target $\pi_{h,t}^{target}$ is publicly stated by the BoK and varies over time $t$.

The key difference between our approach and that of \citet{granziera2025bias} is that the BoK explicitly states its inflation target, while the ECB aims to keep inflation “below, but close to, 2\%.” Consequently, the ECB’s inflation target must be inferred, while the BoK’s target is known and directly incorporated into our analysis.

\citet{granziera2025bias}, adopt the method proposed by \citet{odendahl2023evaluating} to simultaneously estimate the threshold parameter $\pi^*$ and test for zero-mean forecast bias. They estimate a threshold of 1.8\%. In contrast, since the BoK’s threshold is publicly known and time-varying, we directly use this information in our analysis.

Defining the state of the economy for variables other than inflation rate is not straightforward. Several studies have examined the state-dependent bias in output growth rate forecasts using different definitions of economic conditions. For example, \citet{sinclair2010can} use NBER-dated recessions to classify economic states, while \citet{xie2016time} adopt a D-quarter moving average of past growth rates as a proxy. \citet{kpl2011} employ dummy variables based on KOSTAT-dated recessions. However, these definitions can be difficult to justify. In particular, KOSTAT-dated recessions are determined ex post, meaning that the Bank of Korea (BoK) does not have access to this classification at the time forecasts are made.

We extend \citet{holden1990testing} test as follows:
\[e_{h,t} = \alpha_h d_{h,t} + \delta_h (1 - d_{h,t}) + v_{h,t},\]
with the null hypothesis: 
\[H_0: \alpha_h = \delta_h = 0.\]

\begin{table}[t]
	\begin{adjustbox}{width=1\columnwidth,center}
		\centering
\begin{tabular}{ccccc|cccc}
    \multicolumn{9}{c}{State Dependent Holden-Peel Test of CPI Inflation rate Forecast error} \\
    \multicolumn{9}{c}{$e_{h,t}=\alpha_h d_{h,t}+\delta_h(1-d_{h,t}) +v_{h,t}$} \\ \hline\hline
    \multirow{2}{*}{\textbf{Forecast}} & \multirow{2}{*}{\textbf{N}} 
        & \multicolumn{3}{c|}{$H_0: \alpha_h = \delta_h = 0$} 
        & \multicolumn{2}{c}{$H_0^{\alpha_h}: \alpha_h = 0$ vs $H_1^{\alpha_h}: \alpha_h < 0$} 
        & \multicolumn{2}{c}{$H_0^{\delta_h}: \delta_h = 0$ vs $H_1^{\delta_h}: \delta_h > 0$} \\
    & & \textbf{Wald} & \textbf{p-value} &  & \textbf{t} & \textbf{p-value} & \textbf{t} & \textbf{p-value} \\ \hline
    0 & 45 & 3.42  & 0.18 &  & -1.42 & 0.08 & -1.82 & 0.97 \\
    1 & 36 & 20.93 & \textbf{0.00} &  & -3.12 & \textbf{0.00} & 1.69 & \textbf{0.05} \\
    2 & 42 & 5.03  & 0.08 &  & -1.79 & \textbf{0.04} & 0.06 & 0.48 \\
    3 & 25 & 24.92 & \textbf{0.00} &  & -2.03 & \textbf{0.02} & 1.89 & \textbf{0.03} \\
    4 & 26 & 0.76  & 0.68 &  & 0.05 & 0.52 & 0.70 & 0.24 \\ \hline
\end{tabular}

	\end{adjustbox}
\caption{Results of State-Dependent Holden-Peel Test. The test statistics are calculated with the HAC standard errors.}
	\label{tbl:sd_mean_zero}
\end{table}

The null hypothesis implies that the average forecast error is zero, regardless of whether inflation is above or below the target. The first three columns in Table \ref{tbl:sd_mean_zero} report the results of this test. They present the number of observations, the Wald test statistics and their p-values. We reject the null hypothesis $H_0:\alpha_h=\delta_h=0$ for forecast horizons $h=1$ and $h=3$, indicating that for these horizons, the average forecast error differs from zero in at least one state. 

The remaining columns in Table \ref{tbl:sd_mean_zero} present the results of one sided t tests: for the coefficients $\alpha_h$, we test $H_0^{\alpha_h}: \alpha_h = 0$ against $H_1^{\alpha_h}:\alpha_h<0$; for $\delta_h$ coefficients we test $H_0:\alpha_h=\delta_h=0$ against the alternative hypotheses $H_1^{\delta_h}:\delta_h>0$. We test the hypotheses under significance level of 0.05. If rejected, the former implies that the BoK overpredicts inflation rate when actual inflation is below the inflation target at the time of forecast. Similarly, the latter implies that the BoK underpredicts inflation rate if inflation is above the target. For horizons $h=1,2$ and $3$, the results show statistically significant evidence that when inflation is below the target $d_{h,t}=1$, the forecast bias is negative. On the other hand, we find that when the inflation is above the target $d_{h,t}=0$, the test is rejected for forecast horizons $h=1,3$.

In addition to the state-dependent Holden-Peel test, we estimate the following state-dependent version of the Mincer-Zarnowitz equation:
\[y_t = d_{h,t} \times (\alpha_h + \beta_h y_{h,t}) + (1 - d_{h,t}) \times (\gamma_h + \delta_h y_{h,t}) + u_{h,t},\]
with the null hypothesis:
\[H_0: \alpha_h = \gamma_h = 0,\text{ and}~~ \beta_h = \delta_h = 1.\]
Under $H_0$, the forecast $y_{h,t}$ is conditionally unbiased for $y_t$ across all economic states. If $H_0$ is rejected, it implies that the forecast is biased in at least one of the economic states.

\begin{table}[t]
\begin{adjustbox}{width=1\columnwidth,center}
\begin{tabular}{crlcrlcrlcrl|ccc}
\multicolumn{15}{c}{} \\
\multicolumn{15}{c}{State dependent Mincer-Zarnowitz test of CPI inflation rate Forecast} \\
\multicolumn{15}{c}{$y_t=d_{h,t} \times (\alpha_h+\beta_{h} y_{h,t}) + (1-d_{h,t}) \times (\gamma_{h}+\delta_{h} y_{h,t}) + u_{h,t}$} \\
\multicolumn{15}{c}{$H_0: \alpha_h=\gamma_{h}=0, \beta_{h}=\delta_{h}=1$} \\ \hline\hline
\textbf{Forecast} & \multicolumn{2}{c}{\multirow{2}{*}{$\alpha_h$}} & & \multicolumn{2}{c}{\multirow{2}{*}{$\beta_{h}$}}& & \multicolumn{2}{c}{\multirow{2}{*}{$\gamma_{h}$}} & & \multicolumn{2}{c|}{\multirow{2}{*}{$\delta_{h}$}} &  \multirow{2}{*}{\textbf{Wald}} & \multirow{2}{*}{\textbf{p-value}} & \multirow{2}{*}{\textbf{N}}               \\
\textbf{Horizon}  & \multicolumn{2}{c}{}   &                    & \multicolumn{2}{c}{}     &                      & \multicolumn{2}{c}{} &                           & \multicolumn{2}{c|}{}                            &  &  & \\ \hline
0 & -0.07 & (0.05) && 1.02 & (0.02) && 0.06 & (0.11) && 0.95  & (0.04) & 3.99   & 0.41 & 45 \\
1 & 0.06  & (0.16) && 0.79 & (0.07) && 1.07 & (0.18) && 0.76  & (0.03) & 225.59 & \textbf{0.00} & 36 \\
2 & 0.05  & (0.43) && 0.80 & (0.21) && 0.41 & (0.39) && 0.87  & (0.10) & 14.55  & \textbf{0.01} & 42 \\
3 & 0.81  & (0.54) && 0.28 & (0.28) && 4.61 & (0.50) && -0.27 & (0.15) & 105.16 & \textbf{0.00} & 25 \\
4 & 1.27  & (0.67) && 0.46 & (0.27) && 3.80 & (1.49) && -0.28 & (0.45) & 13.16  & \textbf{0.01} & 26 \\ \hline
\end{tabular}
\end{adjustbox}
\caption{Results of State-Dependent Mincer-Zarnowitz Test. Numbers in parentheses represent the standard errors of the corresponding coefficient estimates. The last column presents the test statistics, their p-values, and the number of observations, respectively.}
\label{tbl:state_dependent}
\end{table}

The results of the state-dependent Mincer-Zarnowitz test are summarized in Table \ref{tbl:state_dependent}. Each column presents the OLS estimates of the regression coefficients, with HAC standard errors in parentheses. The last three columns report the Wald test statistic for $H_0$, its p-value, and the number of observations. The p-values are small for all forecast horizons except $h=0$, indicating that the BoK’s inflation forecasts are biased in at least one of the states for most horizons.

\begin{table}[t]
\centering
\begin{tabular}{cccccccc}
\multicolumn{8}{c}{$y_t=d_{h,t} \times (\alpha_h+\beta_{h} y_{h,t}) + (1-d_{h,t}) \times (\gamma_{h}+\delta_{h} y_{h,t}) + u_{h,t}$} \\ \hline\hline
                   & \multicolumn{3}{c}{$d_{h,t}=1$} && \multicolumn{3}{c}{$d_{h,t}=0$}   \\ 
                   & \multicolumn{3}{c}{$H_0: \alpha_h=0, \beta_{h}=1$}  && \multicolumn{3}{c}{$H_0: \gamma_{h}=0, \delta_{h}=1$}           \\
  \textbf{Forecast}& \multicolumn{1}{c}{\multirow{2}{*}{\textbf{Wald}}} & \multicolumn{1}{c}{\multirow{2}{*}{\textbf{p-value}}} &\multicolumn{1}{c}{\multirow{2}{*}{\textbf{N}}} && \multicolumn{1}{c}{\multirow{2}{*}{\textbf{Wald}}} & \multicolumn{1}{c}{\multirow{2}{*}{\textbf{p-value}}}   &\multirow{2}{*}{\textbf{N}}     \\
  \textbf{Horizon} &        &     &    &&       &      \\ \hline
0 & 2.50  & 0.29 & 26 &  & 3.82   & 0.15 & 19 \\
1 & 37.28 & \textbf{0.00} & 23 &  & 161.52 & \textbf{0.00} & 13 \\
2 & 4.64  & 0.10 & 24 &  & 1.78   & 0.41 & 18 \\
3 & 18.44 & \textbf{0.00} & 15 &  & 88.34  & \textbf{0.00} & 10 \\
4 & 4.27  & 0.12 & 15 &  & 9.30   & \textbf{0.01} & 11       \\ \hline
\end{tabular}
\caption{Mincer-Zarnowitz Test Results. The first three columns correspond to periods when the inflation rate is below the target, while the next three columns correspond to periods when it is above the target.}
\label{tbl:each_state}
\end{table}

To analyze the influence of each state, we present the results of the Mincer-Zarnowitz test for each subsample in Table \ref{tbl:each_state}. Each block of three columns includes the Wald test statistic, its p-value, and the number of observations. The first block corresponds to cases where the inflation rate is below the target, while the second block corresponds to cases where inflation is above the target.

For forecast horizons $h=1$ and $h=3$, we find that the BoK’s inflation forecasts are biased in both states. Additionally, the four-quarters-ahead forecast ($h=4$) is found to be inefficient when the state variable $d_{h,t}=0$ (inflation is above the target). In all other cases, we fail to reject the null hypothesis $H_0$ at a 0.05 significance level.

\section{Bias correction}

The state-dependent hypothesis tests in Section 4 reveal a systematic bias in the BoK’s inflation forecasts. This suggests that economic agents can anticipate forecast errors in advance. Rational agents, therefore, are likely to incorporate such information into their own predictions and decision-making processes. In this section, we evaluate selected bias correction strategies and compare their performance against the BoK’s original forecasts.

We are not the first to explore forecast corrections based on the systematic bias in published forecasts. \citet{arai2014using} applies efficiency tests developed by \citet{patton2012forecast} to enhance the accuracy of the Greenbook forecasts. Using a Mincer-Zarnowitz regression augmented with additional terms that include forecast revisions as regressors, \citet{arai2014using} demonstrates the effectiveness of incorporating such adjustments.\footnote{We also evaluated the Mincer-Zarnowitz bias correction. However, the Mincer-Zarnowitz strategy consistently fails to improve forecast performance, even when accounting for state dependency.} The study evaluates forecast accuracy using both recursive and rolling window RMSEs.

\citet{xie2016time} apply the \citet{nordhaus1987forecasting} regression to correct bias in Taiwan’s government GDP growth forecasts. They augment the regression equation by including additional terms that account for the duration of a particular economic state.

\paragraph{Mean Error (ME) Strategy}

For each period $t$, we compute bias-corrected predictions, denoted by $\hat{y}_{h,t}^j$. A straightforward approach to bias correction involves using the average of the last $w$ forecast errors, where $w$ represents the window size. This strategy assumes that recent forecast errors provide a reliable basis for adjusting future predictions. The adjusted forecast is $\hat{y}^{ME,w}_{h,t} = y_{h,t} + \hat{e}^{ME,w}_{h,t}$ where $\hat{e}^{ME,w}_{h,t} = w^{-1}\sum_{s=t-w}^{t-1} e_{h,s}$.

\paragraph{AR(1) Strategy}

We leverage the serial correlation of forecast errors for bias correction. To reduce out-of-sample variance in the estimates, we exclude the constant term from the AR(1) model\footnote{We also used an AR(1) model with a constant term, but the results indicate better performance without it.}: $e_{h,s}=\alpha_he_{h,s-1}+u_{h,s}$. Then, using past $ w-1$ pairs of $\{(e_{h,s},e_{h,s-1})\}_{s=t-(w-1)}^{t-1}$, we estimate the AR(1) coefficient $\hat{\alpha}_h$ and calculate the fitted value $\hat{e}_{h,t}^{AR(1)}$. Then the adjusted forecast is $\hat{y}^{AR(1)}_{h,t} = y_{h,t} + \hat{e}_{h,t}^{AR(1)}$.\\

Further, we exploit the fact that the BoK tends to overpredict or underpredict inflation depending on the state of the economy. We consider the state-dependent version of the ME and AR(1) strategies.

\paragraph{State-dependent Mean Error (SD-ME) Strategy}

We first estimate the state-dependent version of Holden and Peel (1990) equation: $e_{h,s}=\alpha_hd_{h,s}+\delta_h(1-d_{h,s})+v_{h,s}$. Using past $w$ observations $\{e_{h,s}\}_{s=t-w}^{t-1}$ of forecast error, we estimate the coefficients $\hat\alpha_h$ and $\hat\delta_h$, and predict the forecast error $\hat e_{h,s}^{SD-ME}$. Then adjust the forecast by $\hat y_{h,t}^{SD-ME}=y_{h,t}+\hat e_{h,t}^{SD-ME}$.

\paragraph{State-dependent AR(1) Strategy}

We implement the state-dependent version of the autoregressive strategy. The model is $e_{h,s}=\alpha_h^0e_{h,s-1}+\alpha_h^1d_{h,s}e_{h,s-1}+\epsilon_{h,s}$. Then, using past $w-1$ pairs of $\{(e_{h,s},e_{h,s-1}\}_{s=t-(w-1)}^{t-1}$, we estimate both coefficients $\hat\alpha_h^0$ and $\hat\alpha_h^1$ and predict the $t$-period forecast error by $\hat e_{h,t}=\hat\alpha_h^0e_{h,t-1}+\alpha_h^1d_{h,t}e_{h,t-1}$. The bias corrected forecast is $\hat y_{h,t}^{SD-AR(1)}=y_t+\hat e_{h,t}^{SD-AR(1)}$.\\

When estimating the AR(1) coefficient $\alpha_h$, or the average forecast error, it is crucial to account for the information set available at each time period. To ensure proper adjustments, we outline the timeline of forecast and realized value announcements, clarifying which data points are accessible at the time of adjustment.

\paragraph{Information set for bias correction}

To clarify which observations can be used for bias correction, we revisit Figure \ref{fig:diag}. Bias correction is applied in each period after the publication of the forecast. For example, when $EO_t^{1st}$ is published, the bias correction is applied to its forecasts. At this point, information up to period $t-1$ is available. If using forecast errors, the available dataset would include $\{e_{0,s}, e_{1,s}, e_{2,s}, e_{3,s}, e_{4,s}, e_{5,s}\}_{s=1}^{t-1}$. The same vintage of data applies to $EO_t^{2nd}$. For subsequent \textit{Economic Outlook} issues, such as those published in $t+1$, one additional data point is added to the information set.

The timing of when actual realized values become publicly available matters in justifying that the information up to $t-1$ is indeed observable at the time of forecasting. The actual realizations of CPI are typically announced in the month following the end of each quarter. Therefore, we can observe $e_{h,t-1}$ before the publication of $EO_t^{1st}$. For instance, $EO_t^{1st}$ is issued in August and February, while the actual realized inflation rate is announced in July and January, respectively.

%We compute and plot RMSEs for each window sizes. We implement this bias-correction strategy for each fixed forecast horizons. It is useful to check time indices of the available sample at each bias correction. Provided that current time is $t$, for horizons $h=0,1$ we can use up to $t-1$ sample; for $h=2,3$, it is $t-2$, and for $h=4,5$, we can use up to $t-3$ realizations.

%\begin{figure}[ht]
%    \centering
%    \includegraphics[width=1\linewidth]{figures/AR1_optimal_window.png}
%    \caption{RMSE of AR(1) strategy. The black line with circles represent RMSE of AR(1) bias correction strategy for each window size. The red dash-dotted line is RMSE of original forecast without any correction.}
%    \label{fig:RMSE AR(1)}
%\end{figure}

%The result is illustrated in Figure (\ref{fig:RMSE AR(1)}). The black line with circles is the RMSEs of AR(1) bias correction strategy. If AR(1) strategy beats the BoK's forecast --- it is represented by the red dashed line, with a label `No correction' --- the circles are marked with a blue dot. If it fails, then the blue dot is located on zero. For all horizons AR(1) beats BoK, except for $h=3$. The graphs are L-shaped, which implies the recursive window prediction dominates rolling window method. 

\subsection{Results}

To analyze performance of the bias correction strategies, we calculate the ratio of root mean squared forecast errors (RMSFE) for each period $t$. The root mean squared forecast error is estimated by the sample average of the observed out-of-sample squared errors.
\[\widehat{RMSFE}_h^{j}(t) = \sqrt{\frac{1}{|\mathcal T_t|} \sum_{s\in \mathcal T_t} (\hat{y}^{j}_{h,s} - y_s)^2}, \quad j \in \{\text{ME, AR(1), SD-ME, SD-AR(1)}\}, \quad h = 0, 1, 2, 3.\]
where $\mathcal T_t$ denotes the test set at time $t$, $\hat y_{h,s}^j$ denotes bias-corrected forecast value, while $y_s$ denotes the realized value.
Then, the ratio of RMSFE (RRMSFE) of a strategy $j$ in period $t$ with respect to the BoK is calculated as $\widehat{RRMSFE}^{j}(t)=\frac{\widehat{RMSFE}^j(t)}{\widehat{RMSFE}^{BoK}(t)}$. This ratio allows us to compare the relative performance of the bias correction strategies to the BoK's forecasts at each period across different forecast horizons.

The bias-correction strategies are estimated recursively. At each forecast origin $t$, the model is re-estimated with the training set $\{e_{h,s}:s\in \mathcal R_t=\{\text{H2 1999},\dots,t-1\}\}$, and the resulting out-of-sample error is added to the test set. The $\widehat{RMSFE}_h^j(t)$ is then computed recursively from the test set $\{(\hat y^j_{h,s}-y_s)^2 : s \in \mathcal T_t=\{\text{H1 2012},\dots,t\}\}$, which accumulates squared errors starting in H1 2012. This design makes clear the distinction between the training set, which expands with $t$, and the test set, which maintains its starting point.

The number of initial training sample is capped at $|\mathcal R_{\text{H1 2012}}|=25$ semiannual observations (H2 1999--H2 2011), covering the post-crisis recoveries following the Asian Financial Crisis and the Global Financial Crisis. Consequently, when evaluation ends in H1 2016, the number of test observations is $|\mathcal T_{\text{H1 2016}}|=9$; when it ends in H1 2020 (the onset of COVID-19), it is $|\mathcal T_{\text{H1 2020}}|=17$.

This split sizing strikes a balance between model estimation reliability and the need to preserve a sufficiently long hold-out sample; this yields a test-to-train ratio of 0.36 when the evaluation period begins in H1 2016. This period is significant since the BoK has changed its objective from target range with midpoint of $3\%$ to target point of $2\%$ in this period. The test-to-train ratio is consistent with the simulation based guideline by \citet{clark2001tests} of $10\%-40\%$.

Our comparisons focus on the period H1 2016 -- H1 2024, enabling an analysis of performance both before and after the pandemic shock. While this window is long enough to uncover systematic  gains from bias correction, we also highlight an important caveat: direct comparison among different strategies could be misleading. This is because they differ in the pattern and frequency of missing observations.

To accommodate the distinct information requirements of each bias‑correction scheme, we estimate the AR(1), SD‑ME, and SD‑AR(1) adjustments with a recursive expanding window, whereas the model‑free mean‑error (ME) correction is implemented with a rolling window. The logic is straightforward: when a econometric model underlies the adjustment, expanding the estimation sample makes the parameter estimates more accurate; in contrast, the ME procedure gains efficiency exclusively through the choice of window length, so allowing that length to vary is essential.

We embed a validation loop into the ME correction strategy. At every forecast origin $t$, we first choose the window length $w_t$ that minimizes the validation set RMSE. Here all the historical forecast‑error series $\{e_{h,t-s}:s=t-1,\dots,1\}$ up to $t-1$ serves as a validation set. For each candidate window length $w\in\mathcal W$,\footnote{The set of candidate window lengths is set to be $\mathcal W = \{1,2,\dots,50\}$.} we compute the root mean-squared error (RMSE) of the ME-adjusted forecast within the validation set and select the window $w_t$ that minimizes this criterion:
\[w_t=\arg\min_{w\in\mathcal W}RMSE_h^{ME}(w,t),\quad RMSE_h^{ME}(w,t)=\begin{pmatrix}
    \frac{1}{|\mathcal{V}_t|}\sum_{s\in\mathcal{V}_t}(\hat y_{h,s}^{ME,w}-y_s)^2
\end{pmatrix}^{1/2}\]
where $\mathcal V_t$ denotes the validation set consisting of periods from H2 1999 up to $t-1$.\footnote{Unlike the model-based adjustments, the ME strategy does not involve parameter estimation on a training sample. Instead, past forecast errors themselves serve as the basis for selecting the window length $w_t$. In this sense, the validation set for ME plays the role analogous to a training set in the other strategies. As a result, the validation set for ME strategy coincides with the training set of other strategies: AR(1), SD-AR(1), and SD-ME.}
The optimal $w_t$ is then used to construct the bias‑corrected forecast for period $t$.

\begin{figure}[ht]
    \centering
    \includegraphics[width=1\linewidth]{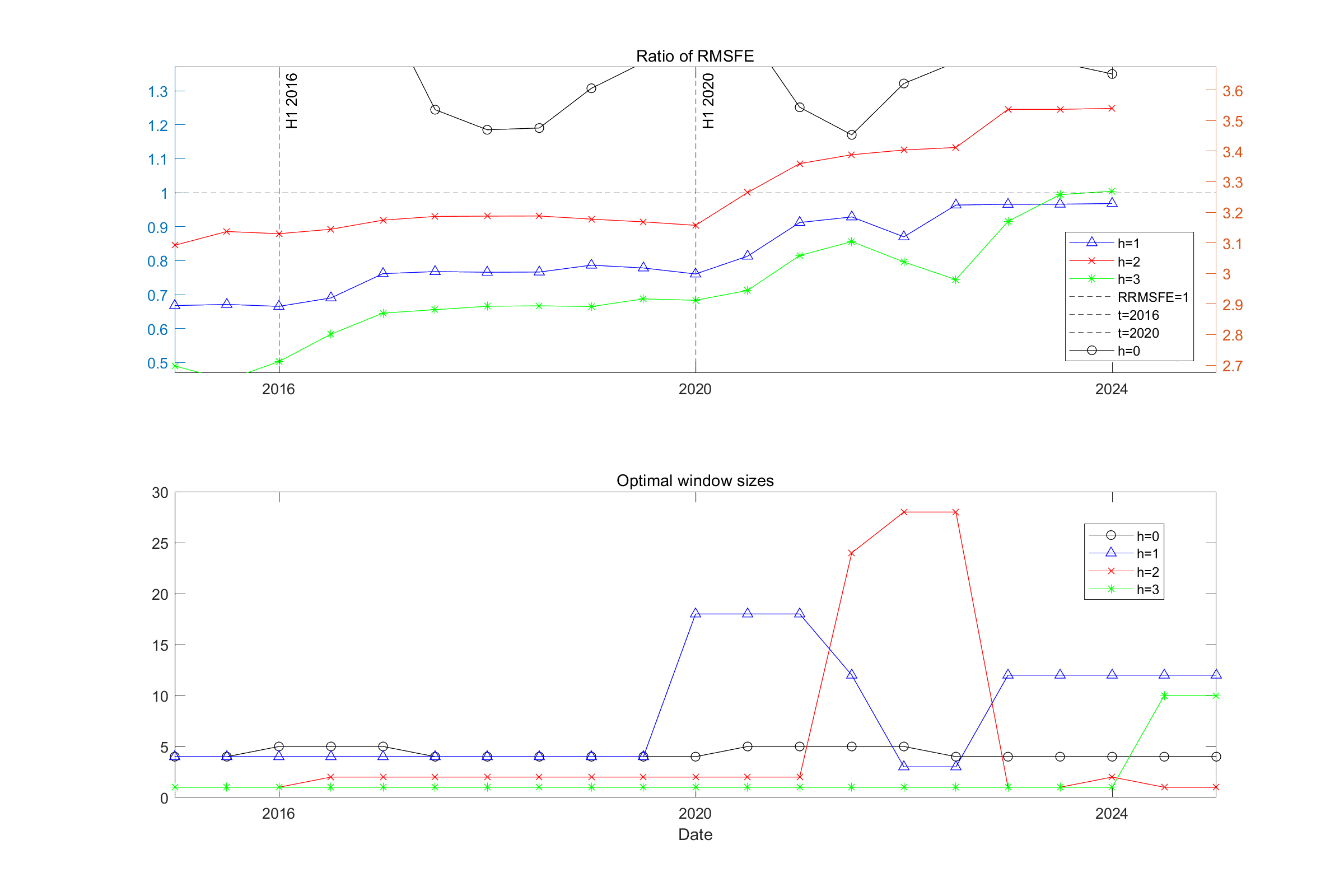}
    \caption{The performance of the ME strategy. The upper panel shows the ratio of RMSFE. The y-axis is the ratio of RMSFE and x-axis is the time period. The scales on the right hand side are designated only for $h=0$. The lower panel shows the path of the chosen window size. The y-axis indicates the window length. $\circ$: $h=0$; $\triangle$: $h=1$; $\times$: $h=2$; $*$: $h=3$.}
    \label{fig:me_strategy}
\end{figure}

Figure \ref{fig:me_strategy} summarizes the outcome of this procedure from H1 2015 through H1 2025. The upper panel plots the ratio of the ME RMSFE to the unadjusted Bank of Korea (BoK) forecast RMSFE. For horizons $h=1,2$, and $h=3$ (blue triangles, red crosses, and green stars, respectively) the ratio falls below unity throughout 2016--2019, indicating sizeable accuracy gains before the COVID‑19 shock; by contrast, the $h=0$ ratio (black circles) hovers near three, reflecting the difficulty of improving upon the BoK’s contemporaneous assessment. After H1 2020, the relative performance deteriorates, and the $h=2$ and $h=3$ ratios eventually rise above one, mirroring the heightened volatility and structural breaks associated with the pandemic period.

The lower panel displays the window length $w_t$ selected at each origin. Two patterns emerge. First, the algorithm overwhelmingly favors $w=4$ for horizons $h=0$ and $h=1$, $w=2$ for horizon $h=2$, and $w=1$ for horizon $h=3$. For short-term forecast horizons ($h=0,1$), structural conditions are fairly stable, so pooling forecast errors over several previous years remains informative. Consequently, the optimal window length is relatively large (around $w=4$). In contrast, for medium-term horizons ($h=2,3$) external shocks frequently change both the sign and magnitude of the bias. As a result, it is optimal to rely only on very recent errors -- typically those from the last two periods ($w=2$).

Second, pronounced spikes appear during and immediately after the COVID‑19 period, with $w_t$ occasionally reaching $25$; these episodes coincide with temporary increases in forecast uncertainty.

%\add{The black line with circles represents the RRMSEs for the AR(1) bias-correction strategy; the green line with stars corresponds to the Mean Error (ME) strategy; the red line with crosses represents the state-dependent AR(1) strategy; and the blue line with triangles corresponds to the state-dependent Mean Error strategy.}

%\add{For horizons $h = 0$ and $h = 2$, the SD-AR(1) strategy outperforms the unconditional AR(1) strategy. In particular, for horizon $h = 2$, its optimal window size is $w = 13$. However, this strategy is not robust: its performance deteriorates significantly for horizons $h = 1$ and $h = 3$. Therefore, we adopt the unconditional AR(1) strategy to ensure consistent performance across all horizons.}

%\add{What is the intuition behind the superior performance of the AR(1) strategy? One possibility is that, as illustrated in Figure \ref{fe_infl}, forecast errors exhibit persistence in both sign and magnitude. Specifically, there is strong positive correlation between $e_{h,s}$ and its lag, $e_{h,s-1}$. Therefore, exploiting this information improves the prediction of forecast errors in the subsequent period.}\footnote{\add{The improvement with a larger window size is straightforward, as the AR(1) coefficient is estimated more precisely with a larger sample. We thank the reviewers for their valuable comments.}}

\begin{figure}[t]
    \centering
    \includegraphics[width=1\linewidth]{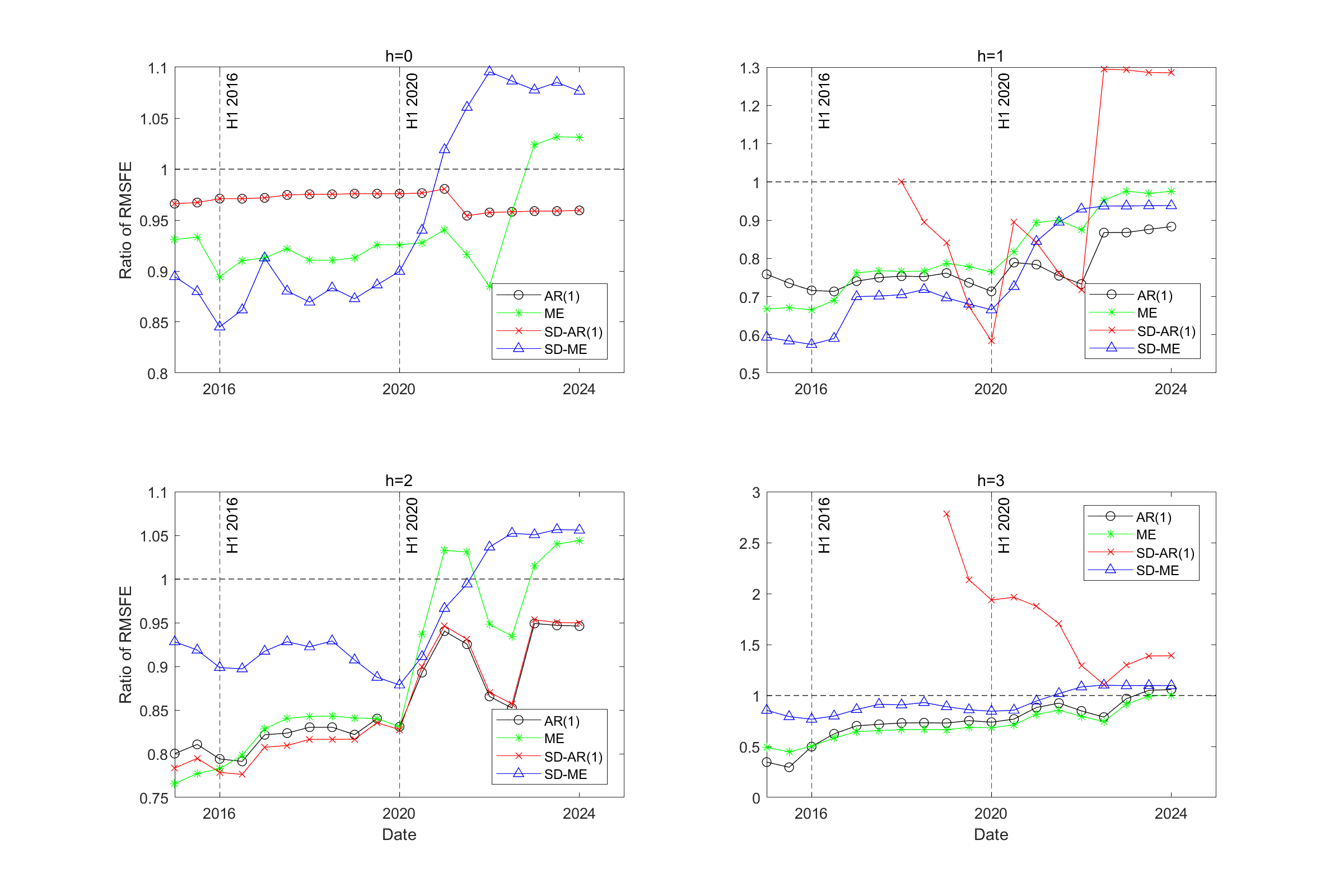}
\caption{The ratios of Root Mean Squared Forecast Errors (RRMSFE) relative to that of the BoK. The x-axis indicates the time period, and the y-axis -- RRMSFE. $\circ$: AR(1); $*$: ME; $\times$: SD-AR(1); $\triangle$: SD-ME.}
    \label{fig:rmse_comparison}
\end{figure}

Figure \ref{fig:rmse_comparison} reports the ratio root‑mean‑squared forecast error (RRMSFE) for each bias‑correction scheme over H1 2015--H1 2024. A value below one signals an improvement on the raw Bank of Korea forecast, whereas values above one indicate deterioration.

The AR(1) adjustment delivers the most robust gains. For every forecast horizon $h \in \{0,1,2,3\}$, the RRMSFE stays below one throughout most of the sample, with particularly strong improvements during the low‑inflation period 2016–2020. An exception arises for horizon $h=3$: after inflation rate peaked in H2 2022, the AR(1) model failed to foresee the subsequent disinflation, producing a marked uptick in RRMSFE.

What is the intuition behind the superior performance of the AR(1) strategy? One possibility is that, as illustrated in Figure \ref{fe_infl}, forecast errors exhibit persistence in both sign and magnitude. Specifically, there is strong positive correlation between $e_{h,s}$ and its lag, $e_{h,s-1}$. Therefore, exploiting this information improves the prediction of forecast errors in the subsequent period.

The SD‑AR(1) variant mirrors the AR(1)’s success for horizons $h=0$ and $h=2$ but degrades noticeably for $h=1$ and $h=3$. Two data‑quality issues drive this pattern. First, BoK publication of multi‑step forecasts was irregular until 2012, leaving gaps that eliminate the lagged‑error observations required to estimate the AR(1) coefficients. Second, because the inflation state exhibits persistence, both states must be observed before the SD‑AR(1) parameters are identifiable. As a result, bias correction for $h=1$ only becomes feasible from 2018 onward, and for $h=3$ from 2019 -- too little data for reliable estimation of RMSE.

Neither the ME nor the SD‑ME strategy improves average accuracy once the evaluation window extends to H1 2024 except for $h=1$. Both schemes do, however, outperform BoK during the tranquil pre‑COVID interval (2016–2019). For one-step-ahead horizon $h=1$, the strategies outperform BoK at all dates 2016-2024. The key limitation is their reliance on simple moving averages of past errors: when the forecast bias flips sign -- as it often does at regime (state) changes -- the ME strategy adjusts sluggishly, especially if the window length exceeds one. Setting the window to one would sharpen the response during transitions but would underperform once the new regime stabilizes, because the BoK rarely repeats the same error magnitude consecutively, that is, the estimated AR(1) coefficient is well below unity.

By construction, the SD‑ME adds a regime split to a Holden–Peel regression; nonetheless, our results show that the state dummy improves bias correction only intermittently, echoing earlier hypothesis‑test evidence that $\alpha_h$ and $\delta_h$ are indistinguishable from zero for $h=0$ and $h=2$.

All four strategies experience a pronounced rise in RRMSFE after H1 2020, coinciding with the COVID‑19 shock. One plausible interpretation is that the systematic component of the BoK's forecast errors shrank relative to an unmodelled uncertainty component driven by extraordinary policy interventions -- most notably the \textit{COVID-19 Emergency Disaster Relief Funds} and the large-scale credit and liquidity programs rolled out for small and medium enterprises (\citet{bianchi2023fiscal}, \citet{koch2024we}). 

\begin{table}[t]
\setlength{\tabcolsep}{9pt}
\renewcommand{\arraystretch}{1.2}
\begin{adjustbox}{width=0.9\textwidth,center}
\begin{tabular}{ccccccccc}
\toprule\toprule
& \multicolumn{4}{c}{\textbf{H1 2016--H2 2019}} & \multicolumn{4}{c}{\textbf{H1 2020--H1 2024}} \\
\cmidrule(lr){2-5} \cmidrule(lr){6-9}
\textbf{Horizon} & AR(1) & ME & SD-AR(1) & SD-ME & AR(1) & ME & SD-AR(1) & SD-ME \\
\midrule
0 & 1.000 & 0.905 & 1.000 & 0.905 & 0.926 & 1.215 & 0.926 & 1.380 \\
1 & 0.740 & 1.124 & 0.674 & 0.989 & 1.040 & 1.249 & 1.390 & 1.270 \\
2 & 0.931 & 1.019 & 0.957 & 0.777 & 1.091 & 1.300 & 1.105 & 1.275 \\
3 & 1.225 & 1.169 & 2.136 & 1.049 & 1.229 & 1.250 & 1.224 & 1.293 \\

%\toprule
%& \multicolumn{4}{c}{\textbf{H1 2016--H1 2024}} & \multicolumn{4}{c}{\textbf{Full Sample (2012--2024)}} \\
%\cmidrule(lr){2-5} \cmidrule(lr){6-9}
%\textbf{Horizon} & AR(1) & ME & SD-AR(1) & SD-ME & AR(1) & ME & SD-AR(1) & SD-ME \\
%\midrule
%0 & 0.952 & 1.118 & 0.952 & 1.237 & 0.959 & 1.031 & 0.959 & 1.076 \\
%1 & 0.973 & 1.219 & 1.285 & 1.206 & 0.883 & 0.975 & 1.285 & 0.938 \\
%2 & 1.051 & 1.231 & 1.067 & 1.163 & 0.946 & 1.044 & 0.950 & 1.056 \\
%3 & 1.228 & 1.235 & 1.390 & 1.249 & 1.058 & 1.004 & 1.390 & 1.096 \\
\bottomrule
\end{tabular}
\end{adjustbox}
\caption{Ratio of RMSFE during two different sub-periods: pre-COVID and post-COVID times. Each panel reports AR(1), ME, SD-AR(1), and SD-ME for horizons 0--3. Values below 1 imply improvement.}
\label{tbl:ratio RMSFE}
\end{table}

Table \ref{tbl:ratio RMSFE} reports the ratios of RMSFE during two sub-periods: the pre-COVID period (H1 2016--H2 2019), and the post-COVID period (H1 2020--H1 2024). Each panel presents AR(1), ME, SD-AR(1), and SD-ME for horizons $h=0$--3. Values below one indicate improvement relative to the BoK forecasts.

The results indicate that in the pre-COVID period bias correction generally yields improvements, except at horizon $h=3$. The only exception is the ME strategy, which does not deliver improvement. In the post-COVID period, most strategies fail to outperform the BoK, except at $h=0$, where AR(1) and SD-AR(1) achieve RRMSFE values below one. Overall, bias correction provided meaningful gains before COVID-19, but performance deteriorated markedly after the outbreak.

In addition, the RRMSFE for the H1 2012--H2 2019 and H1 2012--H1 2024 periods can be checked in Figure \ref{fig:rmse_comparison}. In particular, H2 2019 on the x-axis corresponds to the H1 2012--H2 2019 RRMSFE, and H1 2024 corresponds to the full-sample RRMSFE. Overall, the AR(1) strategy performs consistently compared to the other strategies.

\section{Conclusion}

In this study, we conducted an in-depth analysis of the Bank of Korea’s (BoK) inflation forecasts, uncovering a systematic bias toward the inflation target. Using state-dependent extensions of the Holden and Peel (1990) framework, we demonstrated that the BoK’s inflation forecasts exhibit bias depending on whether inflation is above or below the target. Specifically, the one-quarter-ahead ($h=1$) and three-quarters-ahead ($h=3$) forecasts tend to over-predict inflation when it is below the target and under-predict when it is above. These findings challenge existing conclusions about the BoK’s forecast accuracy and underscore the importance of implementing adjustment strategies to address these biases. Among the bias-correction strategies evaluated, the autoregressive (AR(1)) approach showed promise in improving forecast accuracy, particularly for shorter forecast horizons.

\clearpage 
\appendix

\section*{Appendix: Additional Figure}

\begin{figure}[ht]
	\centering
	\subfloat[GDP Growth rate]{\includegraphics[width=0.9\linewidth]{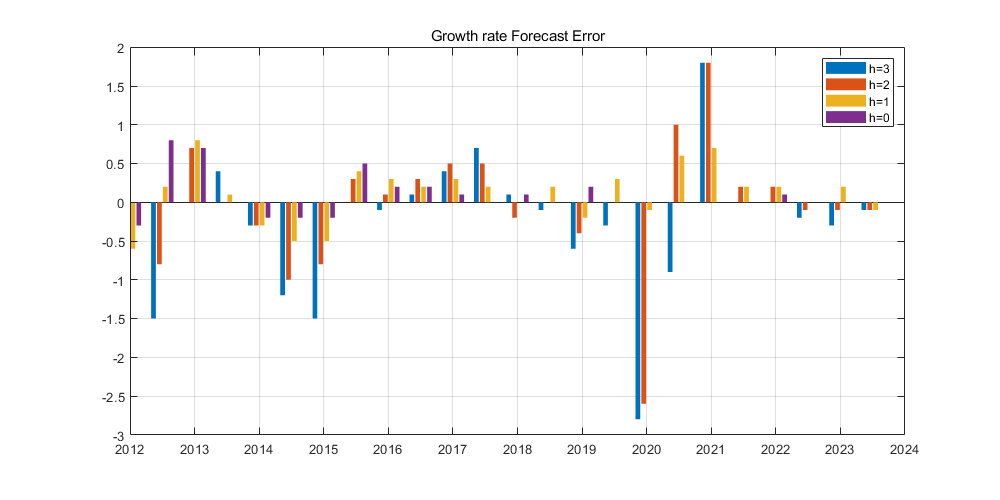}} \\
	\subfloat[Unemployment rate]{\includegraphics[width=0.9\linewidth]{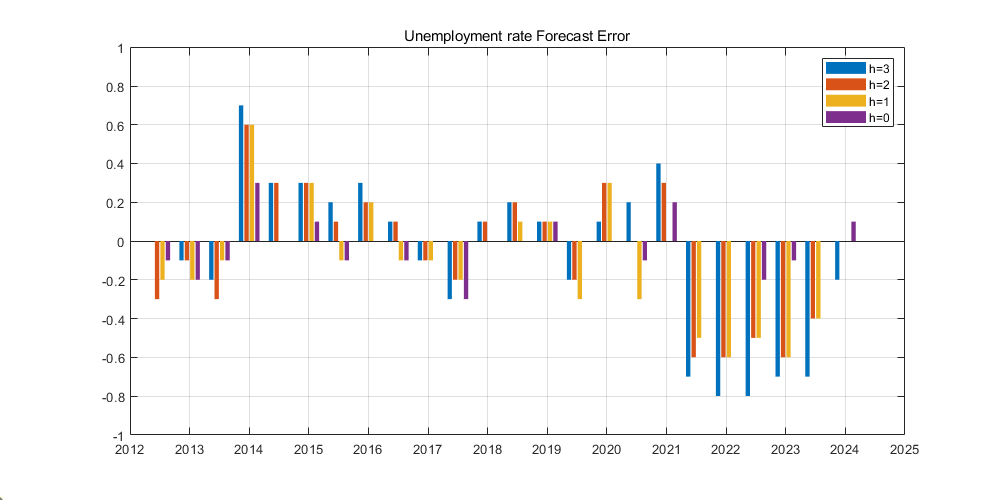}}
	\caption{Forecast Errors for GDP Growth Rate and Unemployment Rate Forecasts}
	\label{fig:forecast_error}
\end{figure}

\newpage

\bibliographystyle{apalike}
\bibliography{ref} % Entries are in the refs.bib file

%\begin{thebibliography}{7}
%    \bibitem{Aldenhoff (2007)} Aldenhoff, Frank-Oliver. 2007. “Are Economic Forecasts of the International Monetary Fund Politically Biased? A Public Choice Analysis.” The Review of International Organizations 2(3): 239–60.

%    \bibitem[2]{survey (2023)} Binder, Carola Conces, and Rodrigo Sekkel. 2023. “Central Bank Forecasting: A Survey.” Journal of Economic Surveys: joes.12554.

%    \bibitem[3]{Timmermann (2007)} Timmermann, Allan. 2007. “An Evaluation of the World Economic Outlook Forecasts.” IMF Staff Papers 54(1): 1–33.

%    \bibitem[4]{Mincer and Zarnowitz (1969)} Mincer, Jacob A., and Victor Zarnowitz. “The Evaluation of Economic Forecasts.” In Economic Forecasts and Expectations: Analysis of Forecasting Behavior and Performance, 3–46. NBER, 1969.
    
%5    \bibitem[5]{kpl2011} 곽노선, 박정수, 이한식. 2011. “국내외 연구기관 전망자료의 예측력 비교 및 평가.” 한국경제연구 29(3): 35–70.

%    \bibitem[6]{ka2015} 김인배, 안희욱. 2015. “중앙은행 커뮤니케이션이 경제주체의 기대형성에 미치는 영향: 한국은행의 GDP 전망공표를 중심으로.” 국제경제연구 21(3): 101–21.

%    \bibitem[7]{sk2008} 손욱, 김영주. 2008. “경제전망의 예측력 및 상호영향력 분석.” 한국경제연구 23: 67–91.
%
%    \bibitem[8]{ck1999} 조장옥, 김준원. 1999. “국내 연구기관 경제전망의 합리성에 관한 분석.” 한국경제의 분석 5(1): 61–104.
%\end{thebibliography}

\end{document}